\documentclass[10pt, twoside, journal]{IEEEtran}
\usepackage{graphics}
\usepackage{amsmath}
\usepackage{latexsym}
\usepackage{amsbsy}
\usepackage{epstopdf}
\usepackage{bm}%
\usepackage{ulem} 
\usepackage{color}

\def\rmA{{\rm A}}
\def\rmB{{\rm B}}

\DeclareFontFamily{U}{cmbsy}{}
\DeclareFontShape{U}{cmbsy}{m}{n}{ <5> <6> <7> <8> <9> gen * cmbsy
       <10> <10.95> <12> <14.4> <17.28> <20.74> <24.88> cmbsy10}{}
\DeclareMathAlphabet{\calb}{U}{cmbsy}{m}{n}

\newcommand{\dyadic}[1]{{#1}
\setbox0=\hbox{$\scriptstyle\leftrightarrow$}
   \setbox2=\hbox{$#1$}
   \dimen0=.5\wd0 \advance\dimen0 by-.5\wd2
   \advance\dimen0 by-\wd0
   \kern\dimen0
{^{\hbox{$\scriptstyle\leftrightarrow$}}}}

\begin{document}

\title{Retrieval Approach for Determining Surface Susceptibilities and Surface Porosities of a Symmetric Metascreen from Reflection and Transmission Coefficients}

\author{~Christopher~L.~Holloway,
        Edward~F.~Kuester, and Abdulaziz H. Haddab
        \thanks{Manuscript received \today.}\thanks{C.L. Holloway, is with the National Institute of Standards and Technology (NIST), U.S. Department of Commerce, Boulder Laboratories,
Boulder,~CO~80305. E.F. Kuester and A.H. Haddab are with the Department of Electrical, Computer and Energy Engineering, University of Colorado, Boulder, CO 80309.  \textcolor{blue}{ christopher.holloway@nist.gov }}}

\markboth{GSTCs for Metasurfaces / Metascreens}{Retrieval for Surface Susceptibilities and Surface Porosities}

\maketitle

\begin{abstract}

Recently we derived generalized sheet transition conditions (GSTCs) for electromagnetic fields at the surface of a metascreen (a metasurface with a ``fishnet'' structure, i.~e., a periodic array of arbitrary spaced apertures in a relatively impenetrable surface).  The parameters in these GSTCs are interpreted as effective surface susceptibilities and surface porosities, which themselves are related to the geometry of the apertures that constitute the metascreen.
In this paper, we use these GSTCs to derive the plane-wave reflection  ($R$) and transmission ($T$) coefficients of a symmetric metascreen, expressed in terms of these surface parameters. From these equations, we develop a retrieval approach for determining the uniquely defined effective surface susceptibilities and surface porosities that characterize the metascreen from measured or simulated data for the $R$ and $T$ coefficients. We present
the retrieved surface parameters for metascreens composed of five different types of apertures (circular holes, square holes, crosses, slots, and a square aperture filled with a high-contrast dielectric). The last example exhibits interesting resonances at frequencies where no resonances exist when the aperture is not filled, which opens up the possibility of designing metasurfaces with unique filtering properties. The retrieved surface parameters are validated by comparing them to other approaches.
\vspace{7mm}

\end{abstract}

\section{Introduction}

A metasurface \cite{hk3} is the surface (or two-dimensional) version of a three-dimensional metamaterial \cite{c1}-\cite{cui}.
The simplicity and relative ease of fabrication of metasurfaces make them attractive alternatives to metamaterials \cite{hk3} and \cite{alex}. We call any periodic two-dimensional structure whose thickness and periodicity are small compared to a wavelength in the surrounding media a metasurface.
We can identify two important subclasses of metasurfaces, characterized by the type of topology possessed by the metasurface.
Metasurfaces that have a ``cermet'' topology, which refers to an array of isolated (non-touching) scatterers, are called metafilms (see Fig.~1a), a term coined in \cite{kmh}. Metasurfaces with a ``fishnet'' structure are called metascreens \cite{metascreen}, see Fig.~1b. In general, metascreens are characterized by periodically spaced apertures in an otherwise relatively impenetrable surface. There are other types of metasurfaces exist
somewhere between these two extremes; for example, a grating of arbitrarily-shaped coated parallel conducting wires called a metagrating. Metagratings behave like a metafilm to electric fields perpendicular to the wire axes and like a metascreen for electric fields parallel to the wire axis \cite{wirehk}. Metafilms have been studied extensively in recent years, but although metascreens have been widely used, relatively less attention has been given to them from an electromagnetic modeling and analysis point of view.

\begin{figure}[t]
\centering
\scalebox{0.55} {\includegraphics*{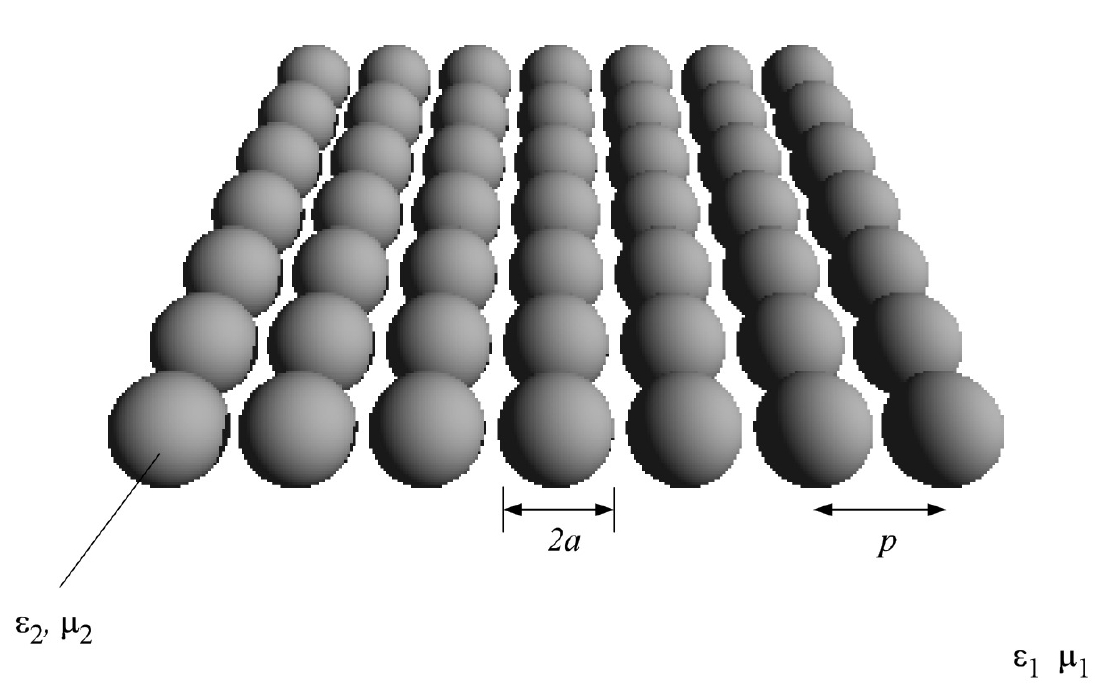}}
\begin{center}
\footnotesize(a) metafilm: special case for an array of spherical particles
\end{center}
\scalebox{0.42} {\includegraphics*{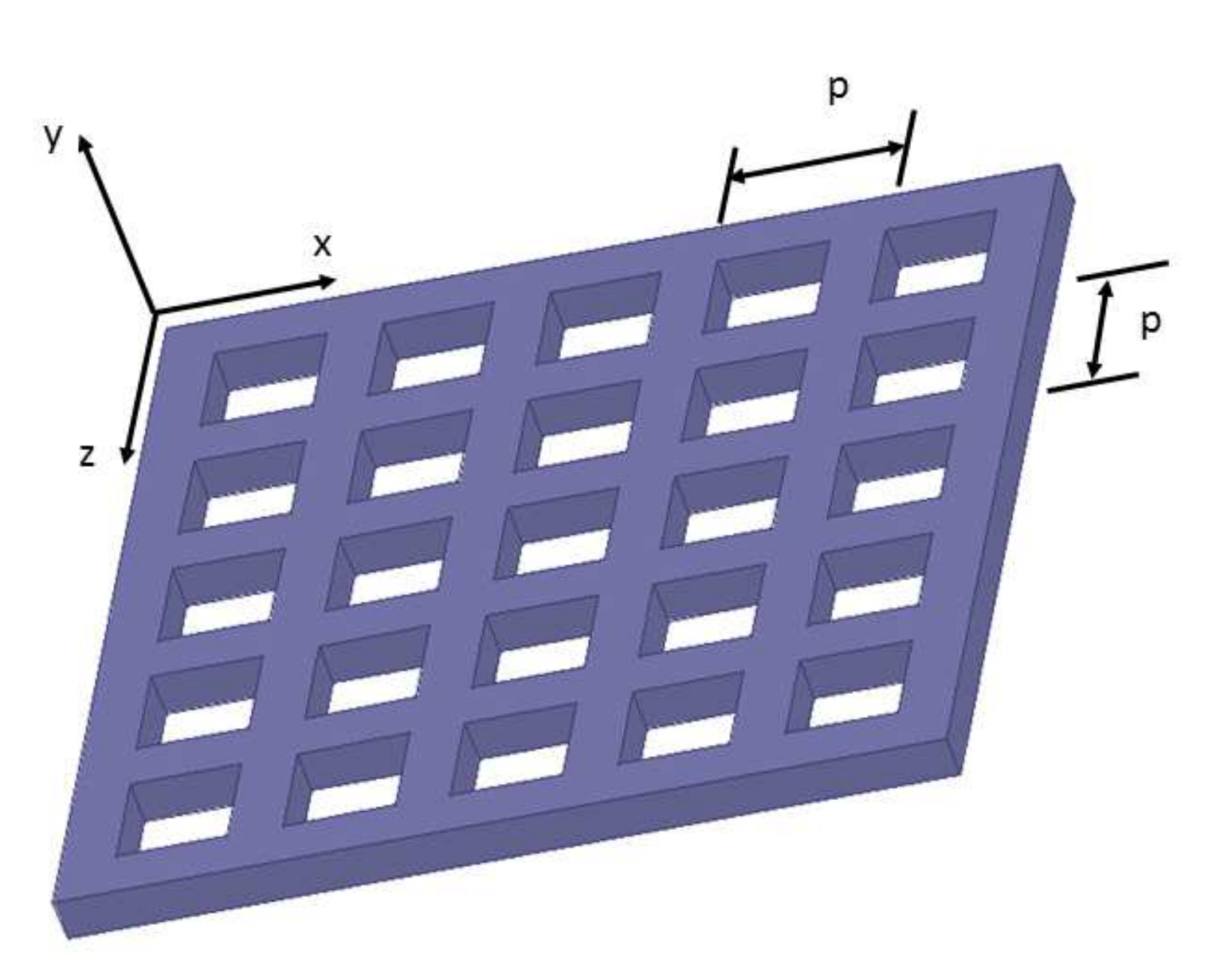}}
\begin{center}
\footnotesize(b) metascreen: special case for an array of square apertures
\end{center}
\caption{Illustration of types of metasurfaces; (a) metafilm which consists of arbitrarily
shaped scatterers placed on the $xz$-plane and (b) metascreen which consists of arbitrarily
shaped apertures in a conducting screen located in the $xz$-plane.}
\label{fig1}
\end{figure}


Like that of a metamaterial, the behavior of a metasurface can be understood in terms of the electric and magnetic polarizabilities of its constituent scatterers (for a metafilm) or apertures (for a metascreen). The traditional and most convenient method by which to model metamaterials is with effective-medium theory, using the bulk electromagnetic parameters $\mu_{\rm eff}$ and $\epsilon_{\rm eff}$. Attempts to use a similar bulk-parameter model for metasurfaces have been less successful, because of ambiguities that arise \cite{hk3}, \cite{hk2}-\cite{hkmetafilm}. In the indicated references, it is shown that the surface susceptibilities of a metafilm are the properties that uniquely characterize a metafilm, and as such, serve as its most appropriate descriptive parameters. As a result, scattering by a metafilm is best characterized by generalized sheet-transition conditions (GSTCs) \cite{kmh}, \cite{hk2}-\cite{hkmetafilm} in contrast to the effective-medium description used for a metamaterial, and the coefficients appearing therein are all that are required to model the macroscopic interaction of a metafilm with an electromagnetic field.  In \cite{hk3}, we stated (without proof) that GSTCs could also be used to model metascreens, and a detailed derivation was given recently in \cite{metascreen} using the method of multiple-scale homogenization. Alternatively, a dipole-interaction model can be used to obtain the GSTCs \cite{ed1} and \cite{ed2}.  However they are obtained, the GSTCs for a metascreen take on a different form than those required for the metafilm, and their features are discussed in \cite{metascreen}, \cite{ed1} and \cite{ed2}.  The issue with a metascreen is that there is the possibility of having tangential surface currents (flowing on the surface of the screen along the $z$ and $x$ directions) that do not vanish as the lattice constant of the metascreen approaches zero, as such, a GSTC that constrains the tangential $H$ cannot be used, see \cite{metascreen} for details.

The GSTCs allow the surface distribution of apertures to be replaced with a boundary condition on the averaged fields that is applied across an infinitely thin equivalent surface (hence the name metascreen), as indicated in Fig. {\ref{fig2}. In \cite{metascreen}, the GSTCs relating the electromagnetic fields on both sides of the metascreen shown in Fig.~\ref{fig1} and Fig.~\ref{fig2} were derived. The size, shape and spacing of the apertures as well as the material properties on both sides of the metascreen make their presence known through effective surface susceptibilities and surface porosities at the interface. It is worth noting that the GSTCs for a metafilm impose conditions on the jumps in the tangential electric and magnetic fields, which depend only on electric and magnetic effective surface susceptibilities. On the other hand, the GSTCs for a metascreen involve one condition for the jump in the tangential electric field, and another for the average of the tangential electric field, the latter of which involves surface porosities. These parameters are uniquely defined and thus represent the physical quantities that uniquely characterize the metascreen. These GSTCs, along with Maxwell's equations, are all that are needed to determine macroscopic scattering, transmission, and reflection from the metascreen.

\begin{figure}
\centering
\scalebox{0.35} {\includegraphics*{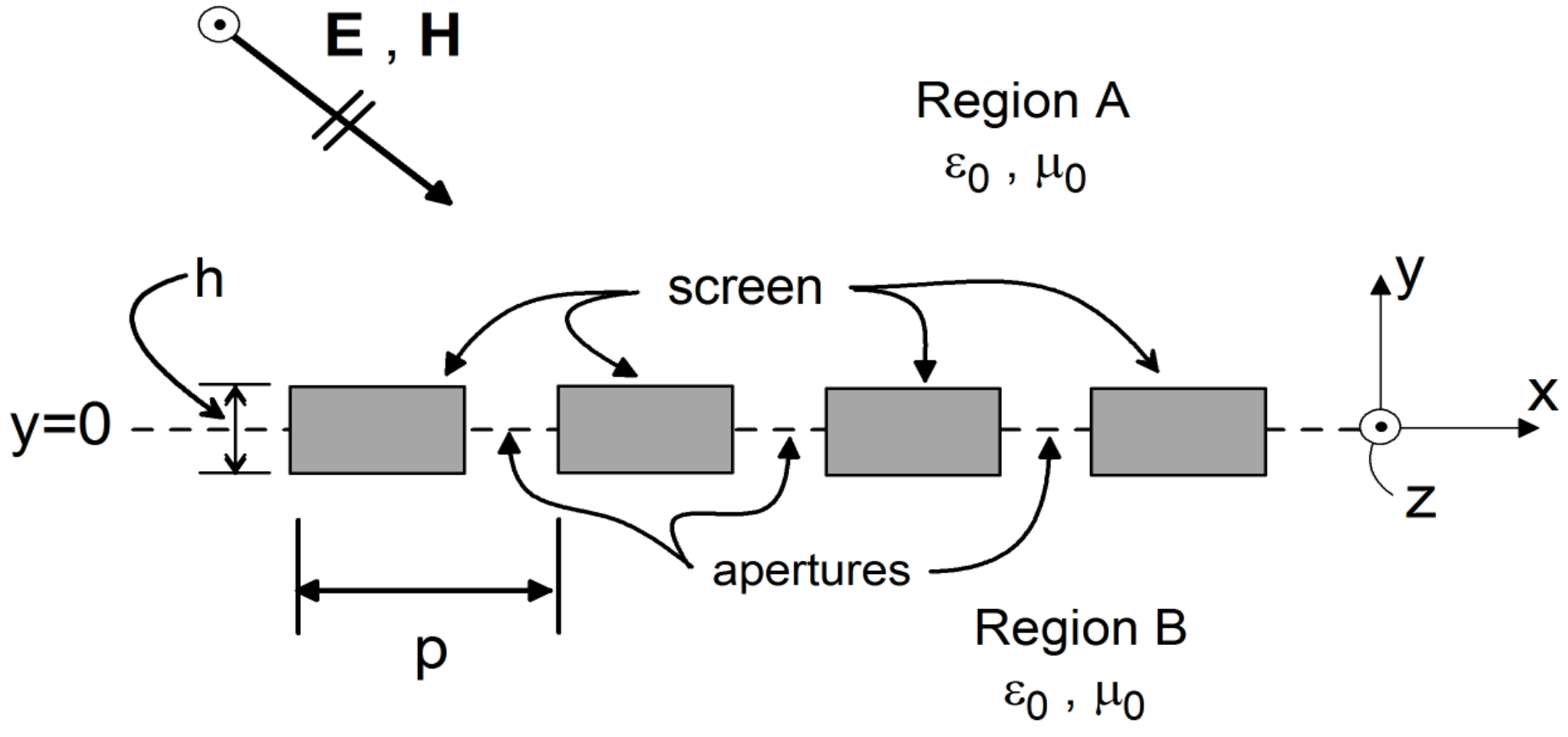}}
\begin{center}
\footnotesize(a)
\vspace{6mm}
\end{center}
\centering
\scalebox{0.35} {\includegraphics*{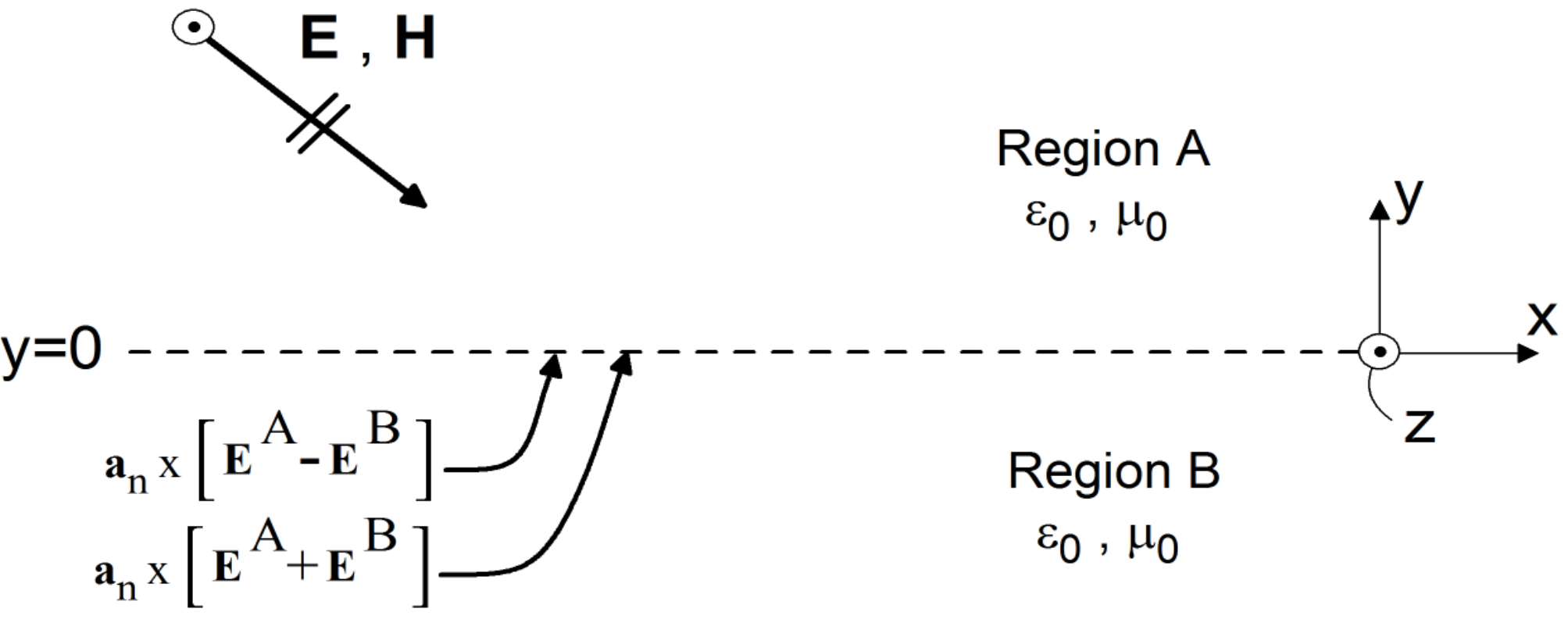}}
\begin{center}
\footnotesize(b)
\end{center}
\caption{(a) Metascreen (array of apertures in conducting screen) and  (b) the reference plane at which the GSTCs are applied.}
\label{fig2}
\end{figure}
\normalsize

In this paper, we use the GSTCs to derive the plane-wave reflection ($R$) and transmission ($T$) coefficients for a metascreen.  The derivation presented in \cite{metascreen} laid out the framework for calculating the required surface susceptibilities and surface porosities, which require the solution of a set of static field problems---this can be computationally challenging for generally shaped apertures.  Thus, we will also derive expressions allowing the surface parameters to be retrieved from measured or simulated values of $R$ and $T$.  This is analogous to the modified Nicolson-Ross-Weir (NRW) approach used for  retrieving the effective permeability and permittivity of a metamaterial \cite{nicolson}-\cite{chen3}, and to the method used to retrieve the surface susceptibilities for a metafilm \cite{hk2} and \cite{awpl}.  Note that the standard NRW approach for metamaterials must be modified when negative material properties exist; typically, the choice of the sign of a square-root is made unambiguous by ensuring an appropriate direction of power flow.  The GSTCs given in \cite{metascreen} will be used here to develop retrieval
techniques for a metascreen.  We demonstrate this retrieval approach by showing results for metascreen composed of an array of five different types of apertures (circular holes, square holes, crosses, slots, and a square aperture filled with a high-contrast dielectric). Finally, we discuss the behavior of the surface parameters (and discuss how to physically interpret them) at the two extreme limits of the aperture fill-factor.

It is worth noting that, while there has been a lot of work on the derivation and use of equivalent boundary conditions (EBC), notably that of Weinstein \cite{vain1, vain} and of Senior and Volakis \cite{senior}, only the EBCs of Sakurai \cite{sak} and of Kontorovich and his colleagues (e.~g., \cite{kont3}) were applicable for metascreen type structures, and these were limited to grids of the thin wire type. Only in \cite{metascreen} have generally applicable GSTCs for a metascreen of fairly general geometry been derived. Of the EBCs discussed in Weinstein \cite{vain1}, eqs. (56.44) and (56.45) therein have the same functional form as those required for a metafilm \cite{kmh}, \cite{hk2}-\cite{hkmetafilm}, while the other set of EBCs [(56.46) and (56.47) therein] are valid for metagrating (wire grating) type structures. However, none of these is valid for a metascreen (the structure analyzed in \cite{metascreen} and the present paper). Only in the particular limit of a metagrating (as discussed in Section IV-D of this paper) does one of our surface parameters reduce to one of Weinstein's surface parameters. There are hints in the book by Senior and Volakis \cite{senior}, as well as in some papers by Volakis and his colleagues \cite{ricoh, topsak}, that a ``universal'' GSTC might exist for an arbitrary thin metasurface, but nowhere does such a condition seem to appear explicitly, and at present we must content ourselves with the observation that topologically distinct metasurfaces (e. g., metafilms, metagratings and metascreens) have distinct GSTCs, none of which is applicable to any of the others, however similar they might appear at a superficial glance. Thus, in order to model and analyze metascreens, the GSTCs presented in this paper are required, on which the techniques for the retrieval of the surface parameters of the metascreen are based.

\section{Reflection and Transmission Coefficients for an Obliquely Incident Plane Wave onto a Symmetric Metascreen}

As seen from the GSTCs derived in \cite{metascreen} for the general case, the dyadic magnetic surface parameters may have off-diagonal terms such that coupling between transverse electric (TE) and transverse magnetic (TM) fields can occur \cite{aniso-metascreen}.  To keep the analysis relatively simple, in this paper we will assume that the apertures are sufficiently symmetric that these off-diagonal terms are zero. We will also assume that regions $A$ and $B$ are both free space, and that the screen possesses mirror symmetry about the reference plane $y=0$. These assumptions apply to many metascreens that are encountered in practice.

Under these conditions, the GSTCs obtained in \cite{metascreen} reduce to:
\begin{equation}
\begin{array}{c}
{\bf{a}}_y\times\left[{\bf{E}}^{\rmA}({\bf r}_o) -{\bf{E}}^{\rmB}({\bf r}_o)\right] = \qquad \qquad \qquad \qquad \\
- j\omega\mu_0 \left[ {\bf{a}}_x \chi_{MS}^{xx} H_{x,{\rm av}}(\mathbf{r}_o) + {\bf{a}}_z \chi_{MS}^{zz} H_{z,{\rm av}}(\mathbf{r}_o) \right] \\
- \chi_{ES}^{yy} {\bf{a}}_y\times \nabla_{t} E_{y,{\rm av}}(\mathbf{r}_o)
\end{array} \,\,\, ,
\label{gstc1b}
\end{equation}
and
\begin{equation}
\begin{array}{c}
{\bf{a}}_y\times{\bf{E}}_{\rm av}({\bf r}_o) = \qquad \qquad \qquad \qquad \qquad \qquad \\
- j\omega\mu_0 \left\{ {\bf{a}}_x \pi_{MS}^{xx} \left[ {H}^{A}_{x}({\mathbf{r}}_o) - {H}^{B}_{x}({\mathbf{r}}_o) \right] \right.\\
+ \left. {\bf{a}}_z \pi_{MS}^{zz} \left[{H}^{A}_{z}({\mathbf{r}}_o) - {H}^{B}_{z}({\mathbf{r}}_o) \right] \right\}\\
- \pi_{ES}^{yy} {\bf{a}}_y\times \nabla_{t} \left[ {E}^{A}_{y}(\mathbf{r}_o) - E^{B}_{y}(\mathbf{r}_o)\right]\\
\end{array}\,\,\, .
\label{gstc2b}
\end{equation}
where
\begin{equation}
 H_{x,{\rm av}}(\mathbf{r}_o) = \frac{1}{2} \left[{H}^{A}_{x}({\mathbf{r}}_o)+{H}^{B}_{x}({\mathbf{r}}_o)\right], \quad \mbox{\rm etc.}
\end{equation}
represent the average of the fields on the two sides of the reference plane at $y=0$. The superscripts $A$ and $B$ correspond to the fields in regions $A$ and $B$, and ${\mathbf{r}}_o$ corresponds to a point in $y=0$, see Fig.~\ref{fig2}. In the foregoing, the surface parameters $\chi_{MS}$ and $\chi_{ES}$  are interpreted as effective electric and magnetic surface {\it susceptibilities} of the metascreen, while the surface parameters $\pi_{ES}$ and $\pi_{MS}$  are interpreted as its effective electric and magnetic surface {\it porosities} \cite{metascreen}. With respect to the general result obtained in \cite{metascreen}, the symmetry of the apertures has resulted in:
\begin{equation}
\pi_{MS}^{xz}=\pi_{MS}^{zx}=\chi_{MS}^{xz}=\chi_{MS}^{zx}\equiv 0\,\,\, ,
\label{sym}
\end{equation}
while the symmetry with respect to either side of the metascreen has given:
\begin{equation}
\begin{array}{rcccl}
\pi_{ES}^{Ayy}=\pi_{ES}^{Byy} \equiv 2 \pi_{ES}^{yy}& ; &
\chi_{ES}^{Ayy}=\chi_{ES}^{Byy} \equiv \frac{1}{2} \chi_{ES}^{yy}\\
\pi_{MS}^{Axx}=\pi_{MS}^{Bxx} \equiv \,\,\,\, 2 \pi_{MS}^{xx}& ; &
\chi_{MS}^{Axx}=\chi_{MS}^{Bxx} \equiv \frac{1}{2} \chi_{MS}^{xx}\\
\pi_{MS}^{Azz}=\pi_{MS}^{Bzz} \equiv \,\,\,\, 2 \pi_{MS}^{zz}& ; &
\chi_{MS}^{Azz}=\chi_{MS}^{Bzz} \equiv \frac{1}{2} \chi_{MS}^{zz}\\
\end{array} \,\, .
\label{symb}
\end{equation}


\subsection{TE Plane Wave Incident on a Metascreen}

Let a metascreen be located in the plane $y=0$ in free space. Assume a TE polarized plane wave is incident onto the metascreen as
shown in Fig.~\ref{fig3}(a), such that the total $E$-field in
region A ($y>0$) is given by ${\bf E} = {\bf E}^i + {\bf E}^{r}$, where the incident and reflected fields are
\begin{equation}
\begin{array}{rcl}
{\bf E}^i & = & {\bf a}_{z} E_{0} e^{-j{\bf k}_{i}{\cdot}{\bf r}} \\
{\bf E}^{r} & = & {\bf a}_{z} R_{TE} E_0
e^{-j{\bf k}_{r}{\cdot}{\bf r}}
\end{array} \,\,\, .
\label{ey1}
\end{equation}
The transmitted field in region B ($y<0$) is given by
\begin{equation}
 {\bf E}^{t}={\bf a}_{z} T_{TE} E_{0} e^{-j{\bf k}_{t}{\cdot}{\bf
 r}}\,\,\, ,
\label{ey2}
\end{equation}
where
\begin{equation}
\begin{array}{rcl}
{\bf k}_{t}= {\bf k}_{i}&=&\left\{{\bf a}_{x}\sin \theta -{\bf a}_{y}
\cos \theta \right\} k_{0}\\
{\bf k}_{r}&=&\left\{{\bf a}_{x}\sin \theta +{\bf a}_{y} \cos
\theta \right\} k_{0}
\end{array}\,\,\, ,
\end{equation}
\begin{equation}
{\bf r}=x{\bf a}_x+y{\bf a}_y++z{\bf a}_z \,\,\, ,
\label{rrrr}
\end{equation}
and $k_0 = \omega \sqrt{\mu_0 \epsilon_0}$ is the wavenumber of
free space. In (\ref{ey1}) and (\ref{ey2}), $R_{TE}$ and
$T_{TE}$ are the reflection and transmission coefficient, respectively.
From Maxwell's equations, the incident, reflected and transmitted
$H$-fields are given by:
\begin{equation}
\begin{array}{rcl}
{\bf H}^i &=&\left(\frac{k_{0}E_{0}}{\omega\mu}\right)\left\{-{\bf a}_{x}\cos \theta -{\bf a}_{y}
\sin \theta \right\} e^{-j{\bf k}_{i}{\cdot}{\bf r}}\\
{\bf H}^{r}&=& \left(\frac{k_{0}E_{0} R_{TE}}{\omega\mu}\right) \left\{{\bf a}_{x}\cos \theta -{\bf a}_{y}
\sin \theta \right\}e^{-j{\bf k}_{r}{\cdot}{\bf r}}\\
{\bf H}^{t}&=& -\left(\frac{k_{0}E_{0}T_{TE}}{\omega\mu}\right) \left\{{\bf a}_{x}\cos \theta +{\bf a}_{y} \sin \theta \right\}e^{-j{\bf k}_{t}{\cdot}{\bf r}}
\end{array} \,\,\, .
\label{hh1}
\end{equation}

\begin{figure}
\centering
\scalebox{0.35} {\includegraphics*{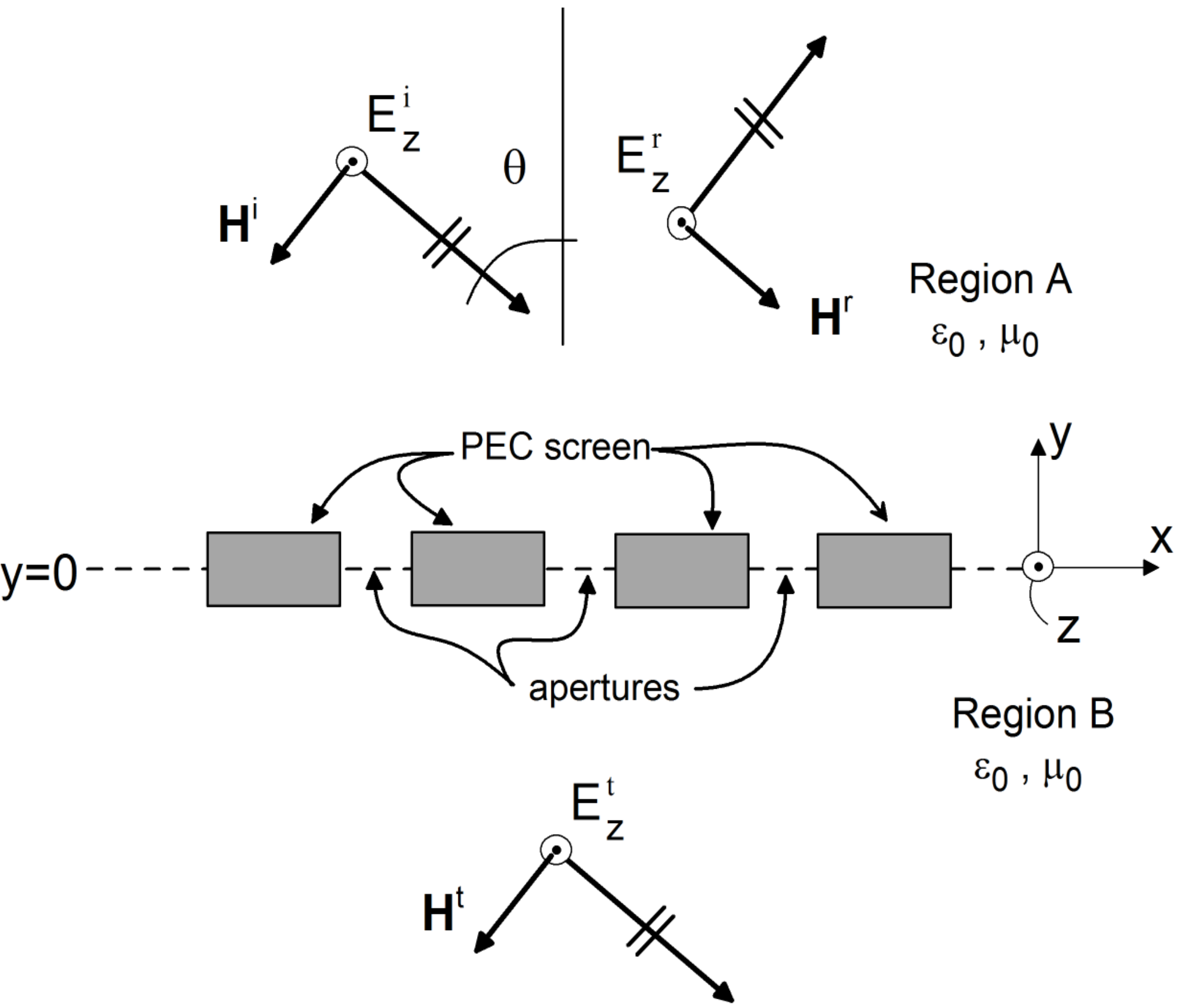}}
\begin{center}
{\footnotesize{(a) TE polarization}}
\vspace{6mm}
\end{center}
\centering
\scalebox{0.35} {\includegraphics*{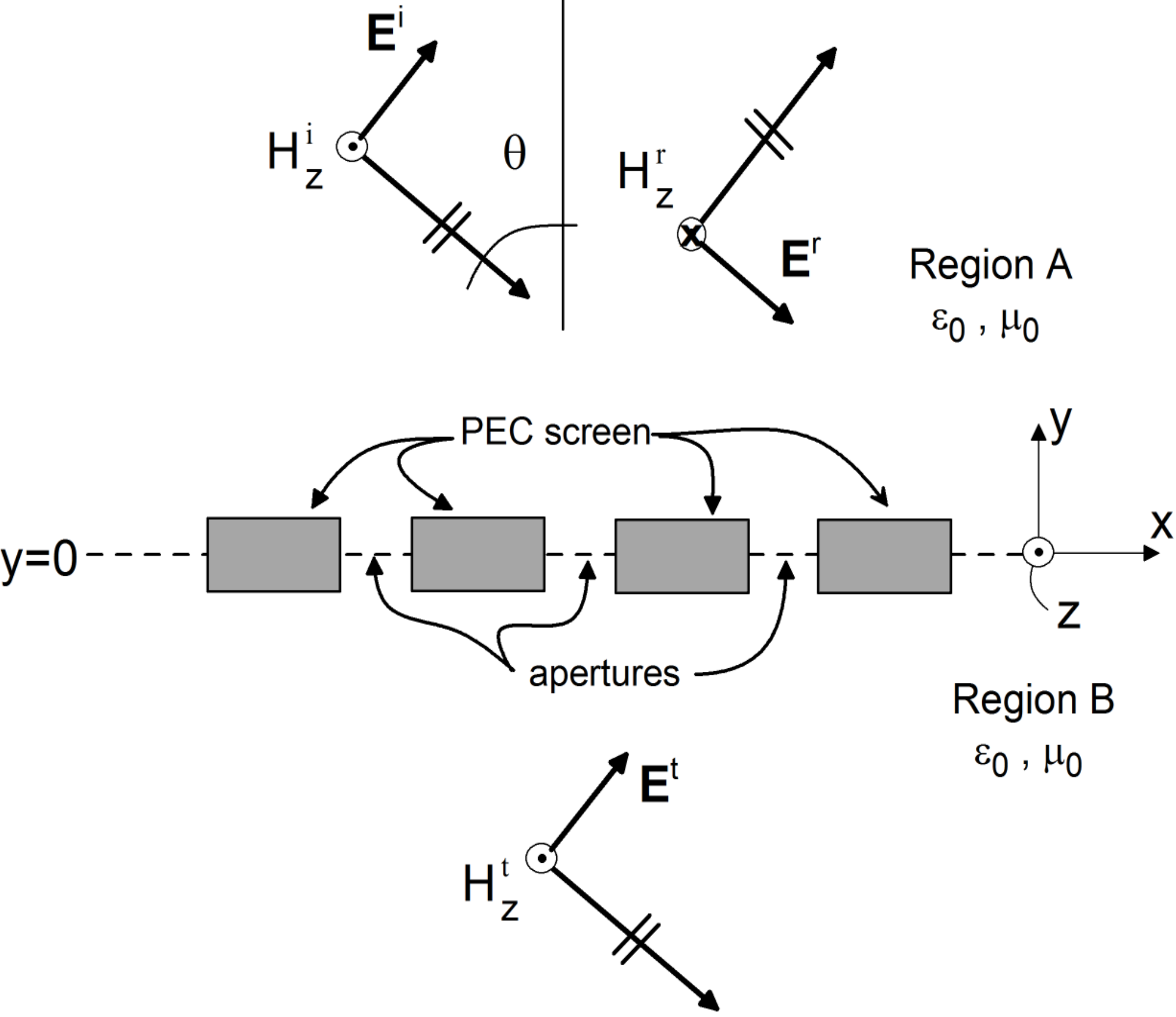}}
\begin{center}
{\footnotesize{(b) TM polarization}}
\end{center}
\caption{Plane wave incident onto a metascreen: (a) TE polarization and (b) TM polarization.}
\label{fig3}
\end{figure}
\normalsize

Substituting the electric and magnetic field components given
in (\ref{ey1}), (\ref{ey2}) and (\ref{hh1}), into the
GSTCs (\ref{gstc1b}) and (\ref{gstc2b}) results in
\begin{equation}
\small
R_{TE}(\theta)=\frac{1+k_0^2\,\chi_{MS}^{xx}\,\pi_{MS}^{xx}\,\cos^2\theta}
{k_o^2\chi_{MS}^{xx}\,\pi_{MS}^{xx}\,\cos^2\theta-i\frac{k_0}{2}
\left(\chi_{MS}^{xx}+4\pi_{MS}^{xx}\right)\cos\theta
-1}
\label{ter}
\end{equation}
and
\begin{equation}
\small
T_{TE}(\theta)=\frac{\frac{i\,k_o\cos\theta}{2}\left[ \chi_{MS}^{xx}-4\pi_{MS}^{xx}\right]}
{k_o^2\chi_{MS}^{xx}\,\pi_{MS}^{xx}\,\cos^2\theta-i\frac{k_0}{2}
\left(\chi_{MS}^{xx}+4\pi_{MS}^{xx}\right)\cos\theta
-1}
\label{tet}
\end{equation}
Note that for this polarization only two surface parameters ($\chi_{MS}^{xx}$ and $\pi_{MS}^{xx}$) are needed to determine $R_{TE}(\theta)$ and $T_{TE}(\theta)$.  This is in contrast to a metafilm, where three different surface parameters are needed to determine $R_{TE}(\theta)$ and $T_{TE}(\theta)$, see \cite{hk3}, \cite{hk2}, \cite{awpl} and \cite{metafilmemc}. The reason for this is that the two GSTCs for a metafilm are essentailly duals of each other---one constraining the jump in tangential $E$, the other tangential $H$. The normal component of surface magnetic susceptibility enters into the equations in the TE case. For a metascreen, we constrain the jump and average of the tangential $E$ field, but tangential $H$ appears only on the right side of the GSTCs multiplied by small factors. Thus, no normal component of surface susceptibility or porosity appears, and one less parameter affects the reflection and transmission.

\subsection{TM Plane Wave Incident on a Metascreen}

Assume a TM polarized $H$-field plane wave is incident onto the
metascreen shown in Fig.~\ref{fig3}(b), such that the $H$-field
components of the incident, reflected, and transmitted plane waves
are given by
\begin{equation}
\begin{array}{rcl}
{\bf H}^{i} &=&
{\bf a}_{z}\frac{E_{0}}{\zeta_0}e^{-j{\bf k}_{i}\cdot{\bf r}} \\
{\bf H}^{r} &=& - {\bf a}_{z} R_{TM}\frac{E_{0}}{\zeta_0}
e^{-j{\bf k}_{r}\cdot{\bf r}} \\
{\bf H}^{t} &=& {\bf a}_{z}T_{TM}\frac{E_{0}}{\zeta_0} e^{-j{\bf
k}_{t}\cdot{\bf r}}
\end{array} \,\,\, .
\end{equation}
where $\zeta_0 = \sqrt{\mu_0/\epsilon_0}$ is the wave impedance of free space. From Maxwell's equations, the incident, reflected and transmitted
$E$-fields are given by:
\begin{equation}
\begin{array}{rcl}
{\bf E}^{i} &=&
E_{0}\left({\bf a}_{x}\cos \theta +{\bf a}_{y}
\sin \theta \right)e^{-j{\bf k}_{i}\cdot{\bf r}} \\
{\bf E}^{r} &=&
E_{0}\left({\bf a}_{x}\cos \theta - {\bf a}_{y}
\sin \theta \right) R_{TM} e^{-j{\bf k}_{r}\cdot{\bf r}} \\
{\bf E}^{t} &=& E_{0}\left({\bf a}_{x}\cos \theta +{\bf a}_{y}\sin
\theta \right) T_{TM}\,\,e^{-j{\bf k}_{t}\cdot{\bf r}}
\end{array}\,\,\, .
\end{equation}
Substituting these sets of expressions into the GSTCs given in (\ref{gstc1b}) and (\ref{gstc2b}) we obtain:
\begin{equation}
\small
R_{TM}(\theta)=
\frac{-1-\frac{k^2_0}{\cos^2\theta}\,{\cal A}\,{\cal B}}
{1-\frac{k_0^2}{\cos^2\theta}{\cal A}{\cal B}+\frac{j\,k_0}{2\cos \theta}\left({\cal A}\,+4\,{\cal B}\right)
}\,\,\, ,
\label{tmr}
\end{equation}
and
\begin{equation}
\small
T_{TM}(\theta)=-
\frac{\frac{j k_0}{\cos \theta}\left[{\cal A}-4{\cal B}\right]}{1-\frac{k_0^2}{\cos^2\theta}{\cal A}{\cal B}+\frac{j\,k_0}{2\cos \theta}\left({\cal A}\,+4\,{\cal B}\right)
}\,\,\, ,
\label{tmt}
\end{equation}
where
\begin{equation}
\begin{array}{c}
{\cal A}=\chi_{MS}^{zz}+\chi_{ES}^{yy}\,\sin^2 \theta\\
{\cal B}=\pi_{MS}^{zz}+\pi_{ES}^{yy}\,\sin^2 \theta
\end{array}
\,\,\, .
\label{defineAB}
\end{equation}
Note that for this polarization four surface parameters ($\chi_{MS}^{zz}$, $\pi_{MS}^{zz}$, $\chi_{ES}^{yy}$, and $\pi_{ES}^{yy}$) are needed to determine $R_{TM}(\theta)$ and $T_{TM}(\theta)$.  This is in contrast to a metafilm, where only three different surface parameters are needed to determine $R_{TM}(\theta)$ and $T_{TM}(\theta)$, see \cite{hk3}, \cite{hk2}, \cite{awpl} and \cite{metafilmemc}.

\section{Retrieval Algorithms for the Surface Parameters}

It is useful to be able to determine the surface parameters that characterize a metascreen by a method other than direct numerical computation as discussed in \cite{metascreen}. Such a retrieval approach for metafilms is presented in \cite{hk3}, \cite{hk2}, and \cite{awpl}. Here we will derive a retrieval approach applicable to metascreens.

\subsection{TE Polarization}

From (\ref{ter}) and (\ref{tet}) it is seen that only two surface parameters ($\chi_{MSS}^{xx}$ and $\pi_{MS}^{xx}$) determine $R_{TE}(\theta)$ and $T_{TE}(\theta)$. Using these two expressions, the two unknown surface parameters are determined from:
\begin{equation}
\pi_{MS}^{xx}=\frac{j}{2k_0}\frac{R_{TE}(0)+T_{TE}(0)+1}{R_{TE}(0)+T_{TE}(0)-1}
\,\,\, ,
\label{pimzz}
\end{equation}
and
\begin{equation}
\chi_{MS}^{xx}=\frac{2j}{k_0}\frac{R_{TE}(0)-T_{TE}(0)+1}{R_{TE}(0)-T_{TE}(0)-1}
\label{chimzz}
\end{equation}
where $R_{TE}(0)$ and $T_{TE}(0)$ are the reflection and transmission coefficients at normal incidence ($\theta=0^{\circ}$). Any angle for $R_{TE}(\theta)$ and $T_{TE}(\theta)$ in (\ref{pimzz}) and (\ref{chimzz}) could have been used, but $\theta=0^{\circ}$ is a convenient choice.  If $R_{TE}(0)$ and $T_{TE}(0)$ are known, either from experimental measurements or from a numerical simulation of the metascreen, then (\ref{pimzz}) and (\ref{chimzz}) can be used to retrieve the values of $\chi_{MSS}^{xx}$ and $\pi_{MS}^{xx}$.
However, if experimental data is used in these retrieval expressions, an angle other than $\theta=0$ may be required because measuring the reflection coefficients at $\theta=0$ can be difficult.  Once these surface parameters are obtained, they can be used to determine $R$ and $T$ for any angle of incidence, which is demonstrated below.

\subsection{TM Polarization}

From (\ref{tmr}) and (\ref{tmt}) we see that $R_{TM}(\theta)$ and $T_{TM}(\theta)$ depend on four of the surface parameters. Thus, unlike the TE polarization case, two sets of $R_{TM}(\theta)$ and $T_{TM}(\theta)$ data are required to determine all four unknowns for the TM polarized wave:
\begin{equation}
\pi_{MS}^{zz}=\frac{j}{2k_0}\frac{R_{TM}(0)+T_{TM}(0)+1}{R_{TM}(0)+T_{TM}(0)-1}\,\,\, 
\label{pimzz2}
\end{equation}
\begin{equation}
\chi_{MS}^{zz}=\frac{2j}{k_0}\frac{R_{TM}(0)-T_{TM}(0)+1}{R_{TM}(0)-T_{TM}(0)-1}\,\,\, 
\label{chimzz2}
\end{equation}
\begin{equation}
\pi_{ES}^{yy}=-\frac{\pi_{MS}^{zz}}{\sin^2 \theta}+\frac{j\cos \theta}{2 k_0\,\sin^2 \theta}\frac{R_{TM}(\theta)+T_{TM}(\theta)+1}{R_{TM}(\theta)+T_{TM}(\theta)-1}\,\,\, 
,
\label{pieyy}
\end{equation}
and
\begin{equation}
\chi_{ES}^{yy}=-\frac{\chi_{MS}^{zz}}{\sin^2 \theta}+\frac{2j\cos \theta}{k_0\,\sin^2 \theta}\frac{R_{TM}(\theta)-T_{TM}(\theta)+1}{R_{TM}(\theta)-T_{TM}(\theta)-1}\,\,\, 
\label{chieyy}
\end{equation}
where $R_{TM}(0)$ and $T_{TM}(0)$ are the reflection and transmission coefficients at normal incidence ($\theta=0^{\circ}$); $R_{TM}(\theta)$ and $T_{TM}(\theta)$ are the reflection and transmission coefficients at some oblique incidence angle, sufficiently different from $\theta=0^{\circ}$. With $R_{TM}(0)$, $T_{TM}(0)$, $R_{TM}(\theta)$ and $T_{TM}(\theta)$ available from either simulation or experiment, (\ref{pimzz2})-(\ref{chieyy}) can be used to retrieve the four unknown surface parameters ($\chi_{MS}^{zz}$, $\pi_{MS}^{zz}$, $\chi_{ES}^{yy}$, and $\pi_{ES}^{yy}$).  Here again, once these surface parameters are obtained, they can be used to determine $R$ and $T$ for any angle of incidence.

In the retrieval approaches for both polarizations, it is important to realize that the reference plane for $R_{TE, TM}(0)$, $T_{TE, TM}(0)$, $R_{TM}(\theta)$ and $T_{TM}(\theta)$ is required to be located at $y=0$. This is a consequence of how the GSTCs were derived in \cite{metascreen}. The GSTCs (and the surface parameters) would need to be modified for different choices of reference plane location \cite{vain}, \cite{senior}, and \cite{hr6}.

It is interesting to observe that the expressions for three of the retrieved surface susceptibilities for a metascreen [$\chi_{MS}^{xx}$, $\chi_{MS}^{zz}$, and $\chi_{ES}^{yy}$, see
eqs.~(\ref{chimzz}), ({\ref{chimzz2}), and (\ref{chieyy})] are the
same expressions as those for three of the surface parameters for a metafilm (see (10) and (11) in \cite{awpl}). There is no reason {\it a priori} to think that this should be the case, and we have only been able to prove this by going through the analysis presented in this paper. The underlying reason for this is most likely that both the metascreen and metafilm have as one of their GSTCs a condition of the same form for the jump in the E-field. We emphasize that a metascreen (treated in this paper) is a very different structure than a metafilm, which was addressed in \cite{awpl}). However, as the second of the GSTCs the metascreen requires a set of conditions on the ``average'' E-fields, while the metafilm has a condition on the ``jump'' in the H-field (the metagrating has GSTCs that are a sort of hybrid of these two conditions). It is worth emphasizing again that the GSTCs required for a metascreen cannot be reduced to those required for a metafilm simply by taking some kind of limit or special case.

Although in this paper we have considered only the case of PEC screen, a similar but more involved derivation can be carried out to derive GSTCs and retrieval algorithms for the case of a screen that is merely a good conductor, and it can be shown the final desired GSTCs will have the same functional form as for the PEC screen, and the retrieval algorithms will have similar forms to those given above. If the screen is not highly conducting (as will be the case for metals at optical frequencies), GSTCs of a rather more complicated form are to be expected---this question is beyond the scope of this paper and will be the topic of a future publication.

\section{Retrieved Surface Parameters for Five Types of Aperture Arrays}

In order to illustrate the validity of these expressions for retrieving the surface susceptibilities and porosities of a metascreen, we will consider five examples.

\subsection{Metascreen Composed of an Array of Circular Apertures}

We first consider an array of circular apertures in a perfect conductor of thickness $h$ as shown in Fig.~\ref{array}(a).  For this metascreen, $p=100$~mm, and $a$ is the radius of the apertures (which will be varied). We show results for three different $h$ ($h=10$~mm, $h=5$~mm, and $h=0.1$~mm). The reflection and transmission coefficients for the metascreen were determined numerically from the finite-element software HFSS (mentioning this product does not imply an endorsement, but serves to clarify the numerical program used) for both $\theta=0^{\circ}$ and $\theta=30^{\circ}$ at a frequency of 500~MHz and $a/p$ ranging from 0 to 0.5.
The numerical values for $R_{TM}(0)$, $T_{TM}(0)$, $R_{TM}(\theta)$ and $T_{TM}(\theta)$ at the reference plane $y=0$ for $h=5$~mm are shown in Fig.~\ref{rtcir}, and were used in (\ref{pimzz2})-(\ref{chieyy}) to determine the four unknown surface parameters ($\chi_{MS}^{zz}$, $\pi_{MS}^{zz}$, $\chi_{ES}^{yy}$, and $\pi_{ES}^{yy}$). Note that since the apertures in the array are symmetric, we have $\pi_{MS}^{xx}=\pi_{MS}^{zz}$ and $\chi_{MS}^{xx}=\chi_{MS}^{zz}$.

\begin{figure}
\centering
\begin{center}
\scalebox{0.25} {\includegraphics*{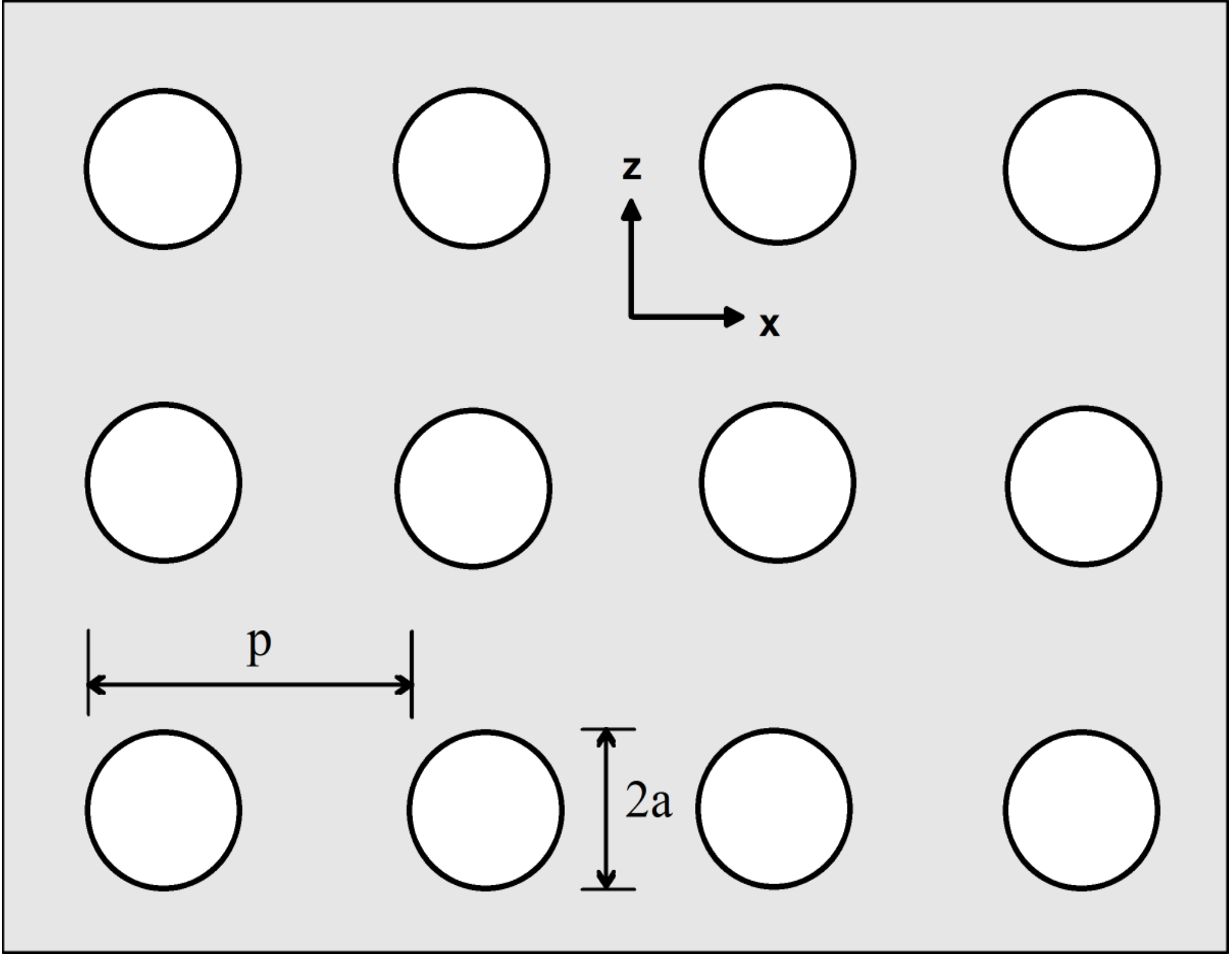}}\\
{\footnotesize{(a) array of circular apertures}}\\
\vspace{6mm}
\scalebox{0.25} {\includegraphics*{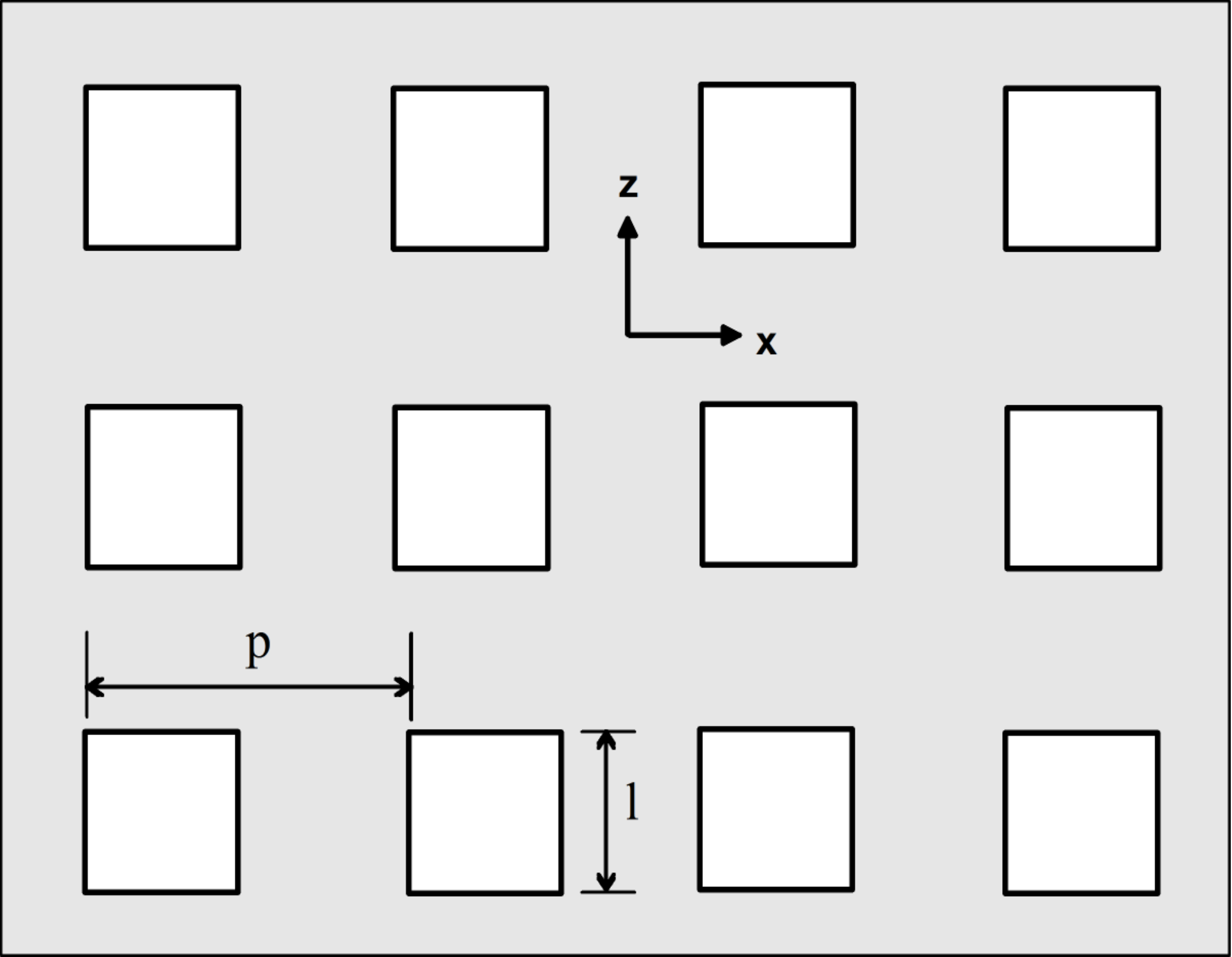}}\\
{\footnotesize{(b) array for square apertures}}\\
\end{center}
\centering
\caption{Metascreen composed of (a) circular apertures and (b) square apertures.}
\label{array}
\end{figure}
\normalsize

\begin{figure}
\centering
\scalebox{0.25} {\includegraphics*{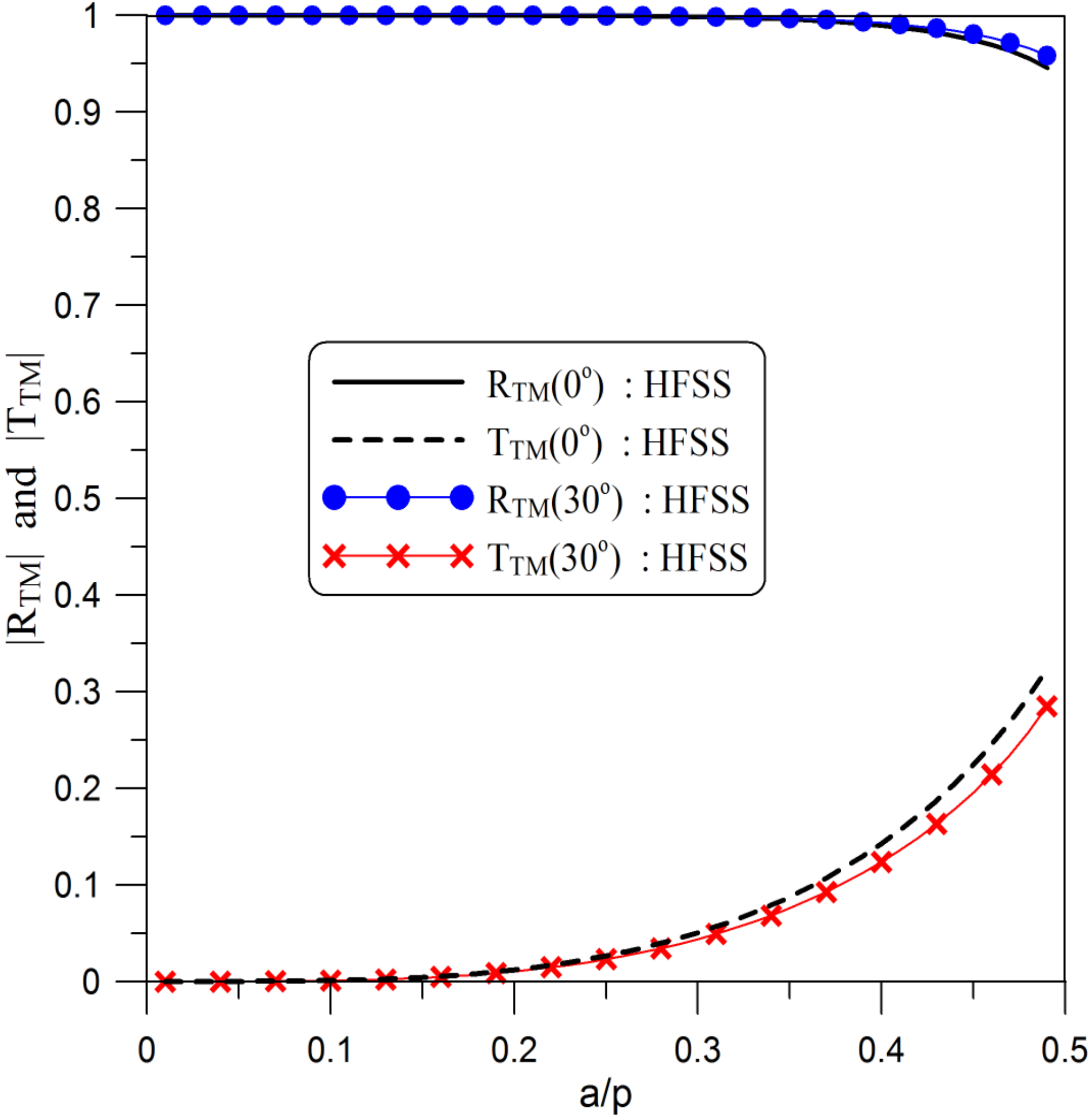}}
\caption{Numerical values for $|R_{TM}|$ and $|T_{TM}|$ for a metascreen composed of circular apertures with $p=100$~mm, $h=5$~mm, and frequency of 500~MHz.}
\label{rtcir}
\end{figure}
\normalsize

Figs.~\ref{pimchim} and \ref{piechie} show the retrieved magnetic and electrical surface parameters for $h=10$~mm, $h=5$~mm, and $h=0.1$~mm.  Using dipole-interaction approximations, closed-form results for these surface porosities and susceptibilities have been obtained in \cite{ed1} for an array of circular apertures (eqs.~(21), (23), (30), (59)-(62) therein). These approximations are also shown in Figs.~\ref{pimchim} and \ref{piechie}.  From the comparison in these two figures, we see that the retrieved surface electric and magnetic susceptibility and surface magnetic porosity correlate very well with the dipole-interaction approximation for the whole range of $a/p=0$ to $a/p=0.49$.  While it is expected that the dipole-interaction approximation may not be valid for closely packed apertures (large $a/p$), it is interesting that good correlation is
observed over the entire $a/p$ range for the array of circular apertures.  The surface parameters can also be determined from a static field solution obtained from an homogenization approach \cite{metascreen}. For further verification that the retrieval approach is valid, we compare to the homogenization approach given in \cite{metascreen}, see Fig.~\ref{hom}.

\begin{figure}
\centering
\scalebox{0.3} {\includegraphics*{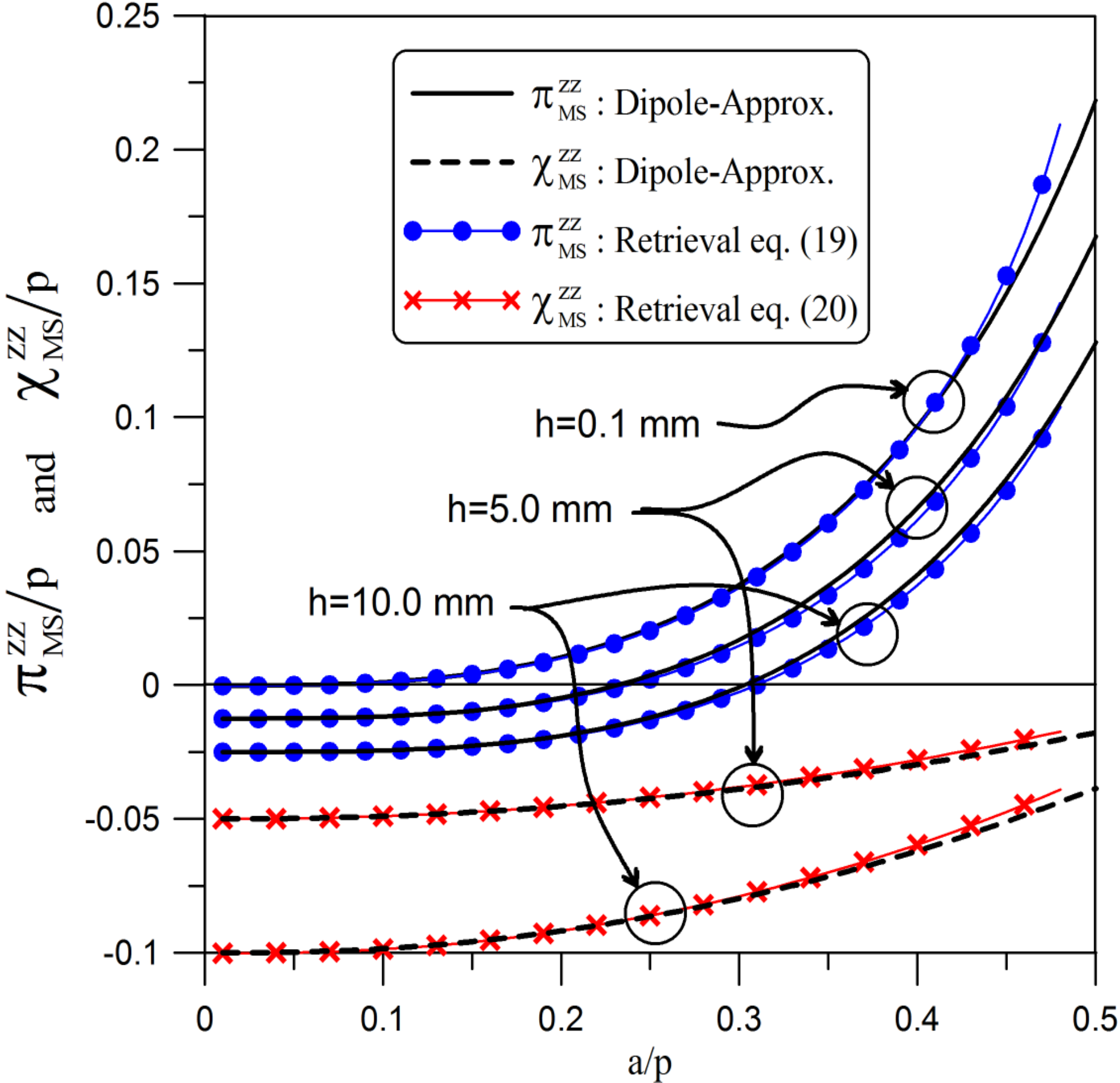}}
\caption{Magnetic surface susceptibility and surface porosity for an array of circular apertures: $p=100$~mm and for $h=10$~mm, $h=5$~mm, and $h=0.1$~mm.}
\label{pimchim}
\end{figure}

\begin{figure}
\centering
\scalebox{0.3} {\includegraphics*{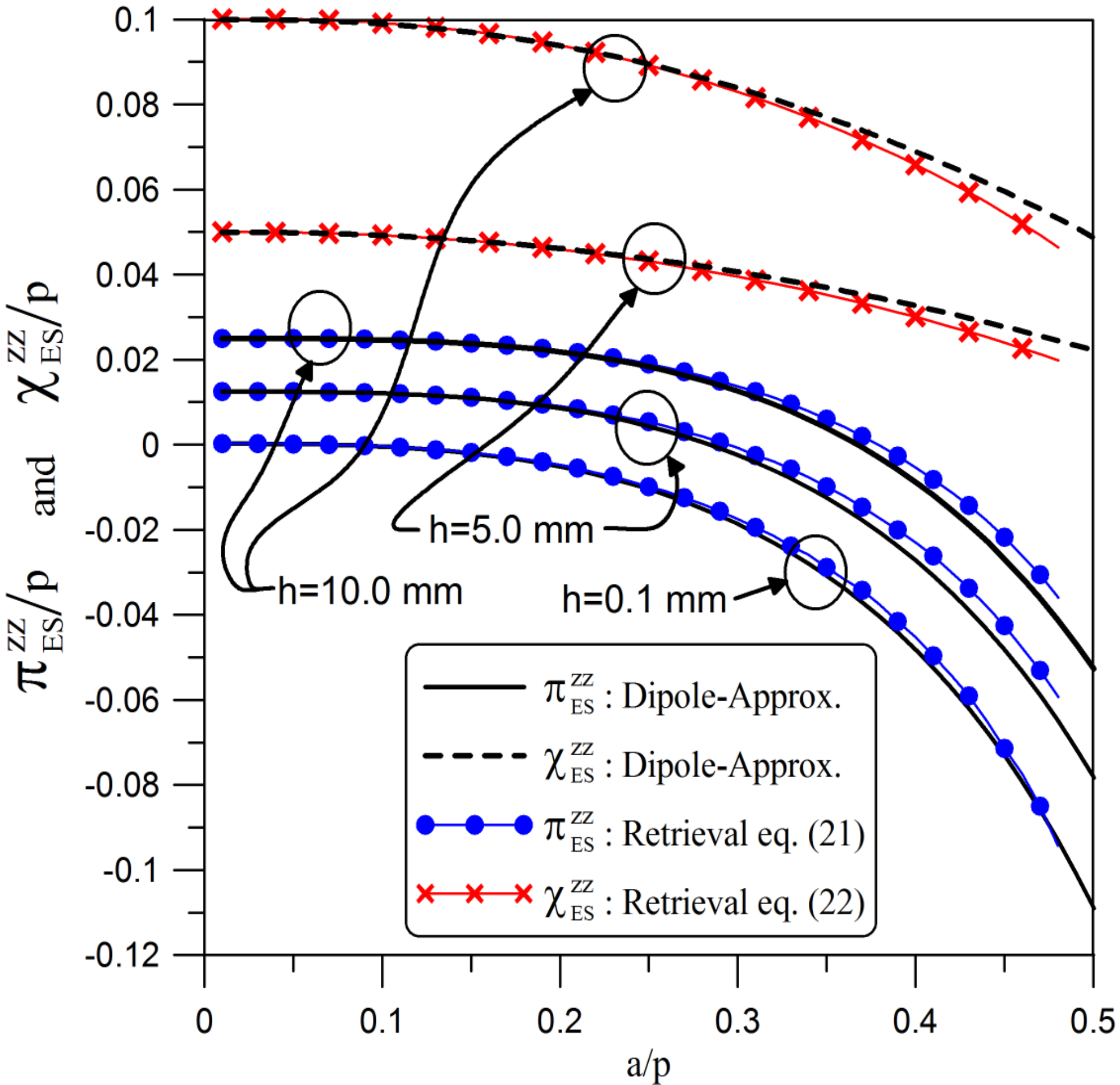}}
\caption{Electric surface susceptibility and surface porosity for an array of circular apertures: $p=100$~mm and for $h=10$~mm, $h=5$~mm, and $h=0.1$~mm.}
\label{piechie}
\end{figure}

\begin{figure}
\centering
\scalebox{0.3} {\includegraphics*{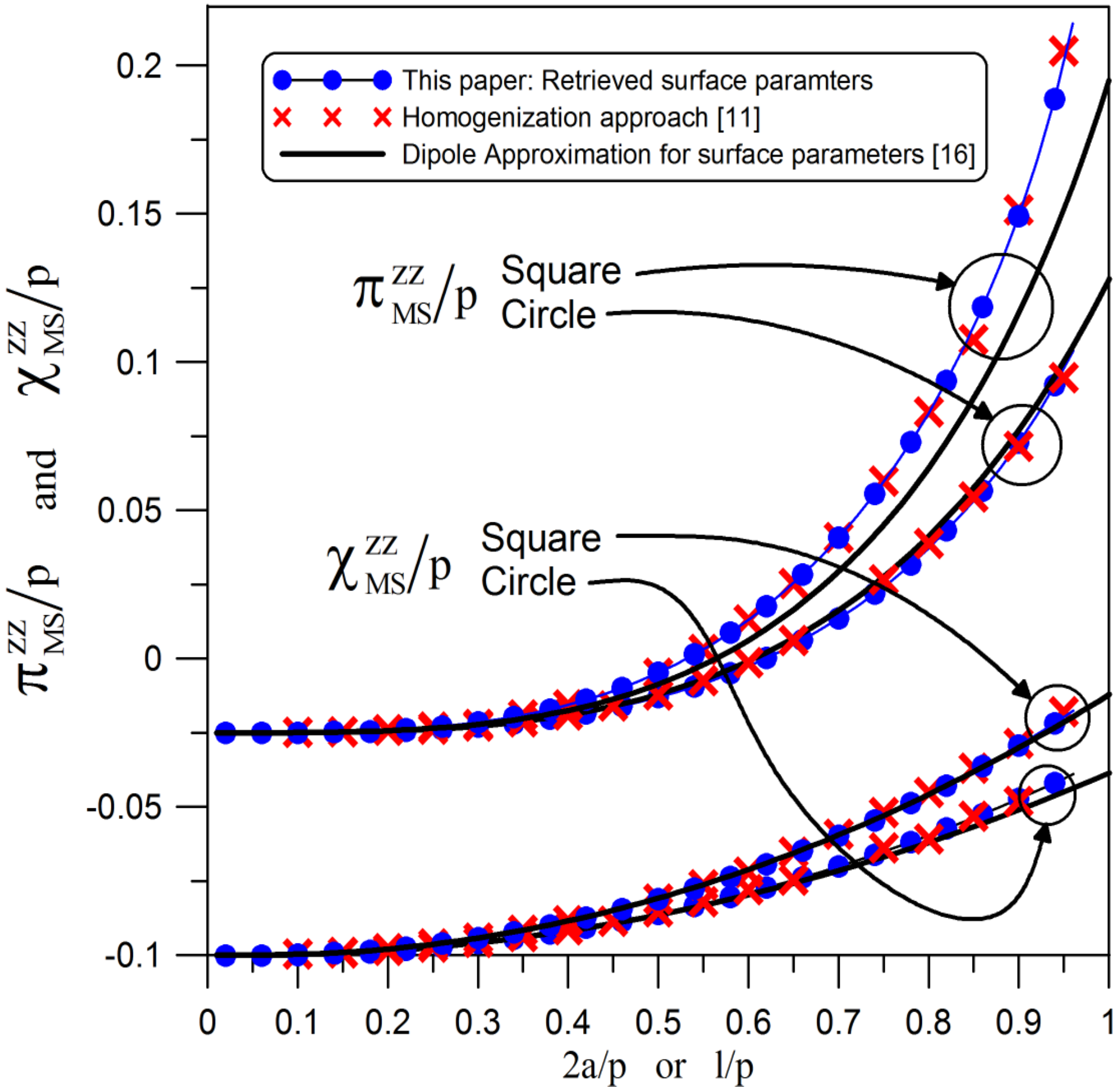}}
\caption{Comparison of the retrieved surface parameters to those obtained from the homogenization approach and to the dipole-approximations.}
\label{hom}
\end{figure}

\subsection{Metascreen Composed of an Array of Square Apertures}

Next, we consider an array of square apertures in a perfectly conducting plate as shown in Fig.~\ref{array}(b).  For this array, $p=100$~mm and $l$ is the length of the side of the square (which will be varied). The reflection and transmission coefficients for the metascreen were again determined numerically with HFSS for both $\theta=0^{\circ}$ and $\theta=30^{\circ}$ at 500~MHz and for $l/p$ ranging from 0 to 1.0.
The numerical values for $R_{TM}(0)$, $T_{TM}(0)$, $R_{TM}(\theta)$ and $T_{TM}(\theta)$ at the reference plane $y=0$ are shown in Fig.~\ref{rtsqr} for $h=5$~mm, and were used in (\ref{pimzz2})-(\ref{chieyy}) to determine the four unknown surface parameters ($\chi_{MS}^{zz}$, $\pi_{MS}^{zz}$, $\chi_{ES}^{yy}$, and $\pi_{ES}^{yy}$). Here again, since the apertures in the array are symmetric, $\pi_{MS}^{xx}=\pi_{MS}^{zz}$ and $\chi_{MS}^{xx}=\chi_{MS}^{zz}$. Figs.~\ref{pimchim2} and \ref{piechie2} show the retrieved magnetic and electrical surface parameters for $h=0.1$~mm, $h=5$~mm and $h=10$~mm.  Approximate
closed-form expressions for these surface parameters for an array of square apertures are given in \cite{ed1} (eqs.~(21), (23), (30), (53)-(66) therein).  These dipole-approximations are also shown in Figs.~\ref{pimchim2} and \ref{piechie2}. From this comparison, the retrieved values for the surface parameters correlated very well to those from the dipole-approximation for $l/p<0.7$.   As discussed below, as $l/p\rightarrow 1$, $\pi_{ES}$ and $\pi_{MS}$ should approach $\infty$. While the retrieved values approach this limit, we see that the dipole-model does not approach this limit. The discrepancy in the dipole-model is most likely due to the fact that some of parameters used in the dipole-approximation for the square aperture give in \cite{ed1} are known to be in error by as much as 20~$\%$ \cite{ed1}. For further verification that the retrieval approach is valid, we compare to the homogenization approach given in \cite{metascreen}, see Fig.~\ref{hom}. This comparison illustrates some of the limitations in the dipole-approximations and indicate that the surface parameters are better determined either from the homogenization-based approach or from the retrieval approach.


\begin{figure}
\centering
\scalebox{0.25} {\includegraphics*{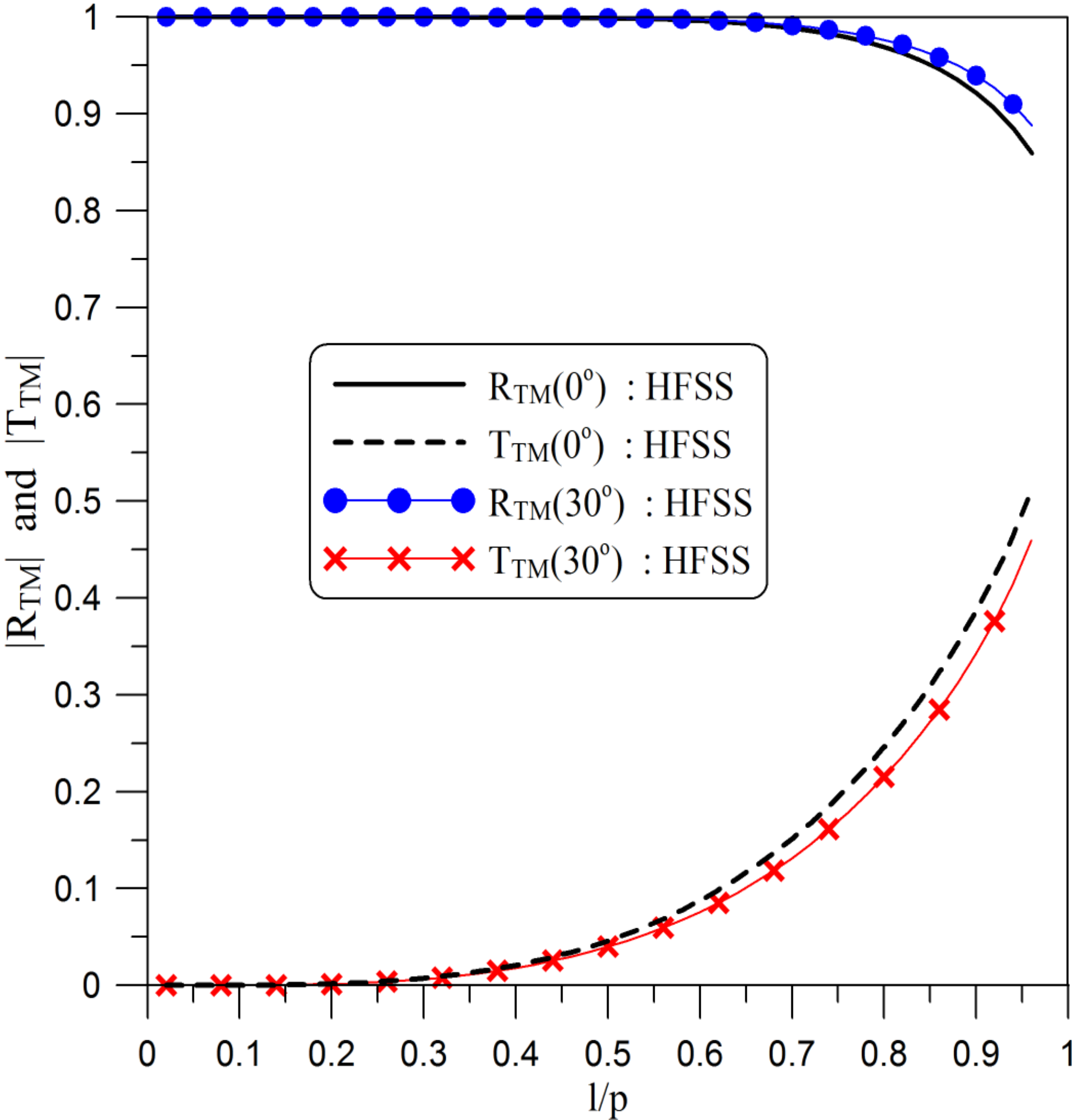}}
\caption{Numerical values for $|R_{TM}|$ and $|T_{TM}|$ for a metascreen composed of square apertures with $p=100$~mm, $h=5$~mm, and frequency of 500~MHz.}
\label{rtsqr}
\end{figure}
\normalsize

\begin{figure}
\centering
\scalebox{0.3}{\includegraphics*{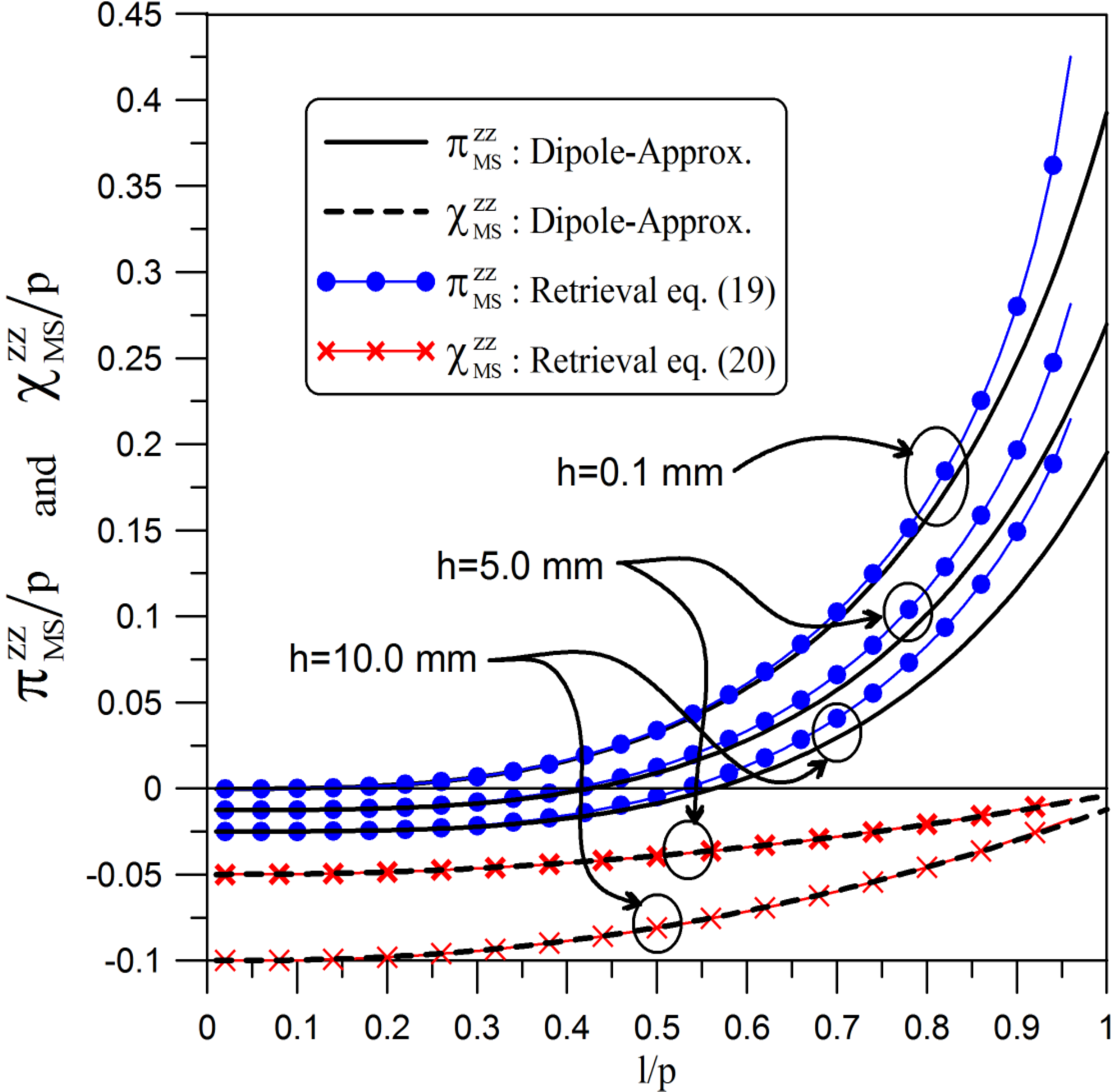}}
\caption{Magnetic surface susceptibility and surface porosity for an array of square apertures: $p=100$~mm and for $h=10$~mm, $h=5$~mm, and $h=0.1$~mm.}
\label{pimchim2}
\end{figure}

\begin{figure}[h!]
\centering
\scalebox{0.3}{\includegraphics*{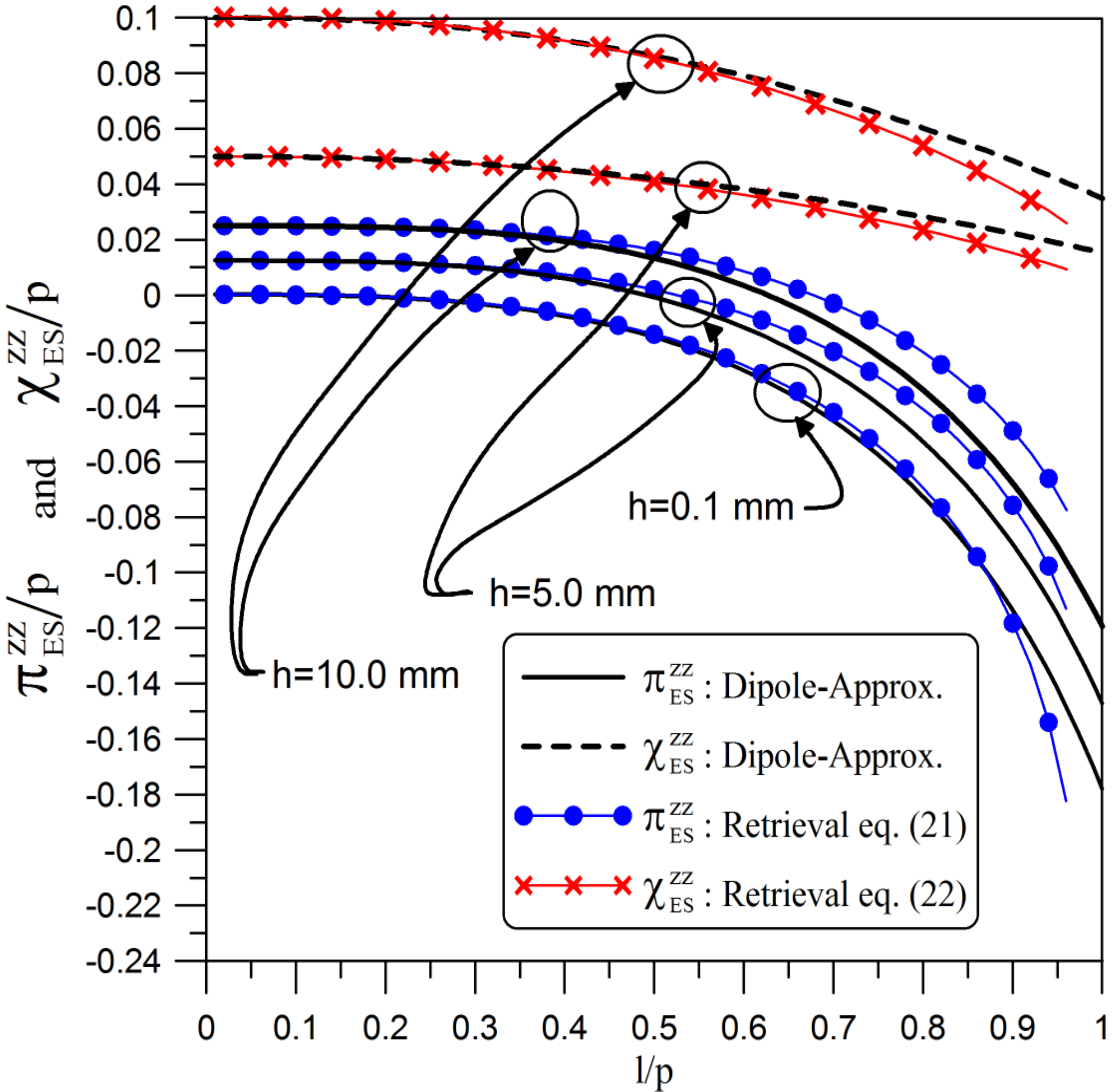}}
\caption{Electric surface susceptibility and surface porosity for an array of square apertures: $p=100$~mm and for $h=10$~mm, $h=5$~mm, and $h=0.1$~mm.}
\label{piechie2}
\end{figure}

\subsection{Metascreen Composed of an Array of Cross-Shaped Apertures}

In this example we consider a metascreen composed of the cross-shape apertures shown in Fig.~\ref{cross}(a).  Here we show results for the surface parameters for various values of the slot length ($l$) and the slot width ($s$). In doing such, we consider the following two cases, (1) we set the length of the crosses to a constant and vary the slot width, and (2) we set the width of the crosses to a constant and vary the slot length.  In all these cases the thickness of the apertures is 5~mm.  Using HFSS we calculate the the $R(\theta)$ and $T(\theta)$ for both $0^{\circ}$ and $30^{\circ}$ incident angles for both the TE and TM polarizations. The reflection and transmission coefficient where used in eqs.~(\ref{pimzz2})-(\ref{chieyy}) to obtained the surface parameters. Fig.~\ref{cross}(b)-\ref{cross}(e) shows results for the surface parameters as a function $s$ for different values for $l$. Note that $\pi_{MS}^{xx}=\pi_{MS}^{zz}$ and $\chi_{MS}^{xx}=\chi_{MS}^{zz}$. Once $s=l$, the aperture reduces to a square aperture, therefore, for a reference, we also show the results of the square aperture.
Fig.~\ref{cross1}(a)-\ref{cross1}(d) shows results for the surface parameters as a function $l$ for different values for $s$.  Again, when $l=s$, the aperture reduces to a square aperture, therefore, for a reference, we also shown the results of the square aperture.

It is interesting to obverse that when $l/p=1$, the metascreen converts to a metafilm (i.e., the array of crosses convert to an array of square plates). When this occurs, one must use the GSTCs for a metafilm (\cite{hk3}, \cite{kmh}, \cite{hk2}-\cite{hkmetafilm}) to analyze the metasurface.

\begin{figure*}
\centering
\scalebox{.2}{\includegraphics{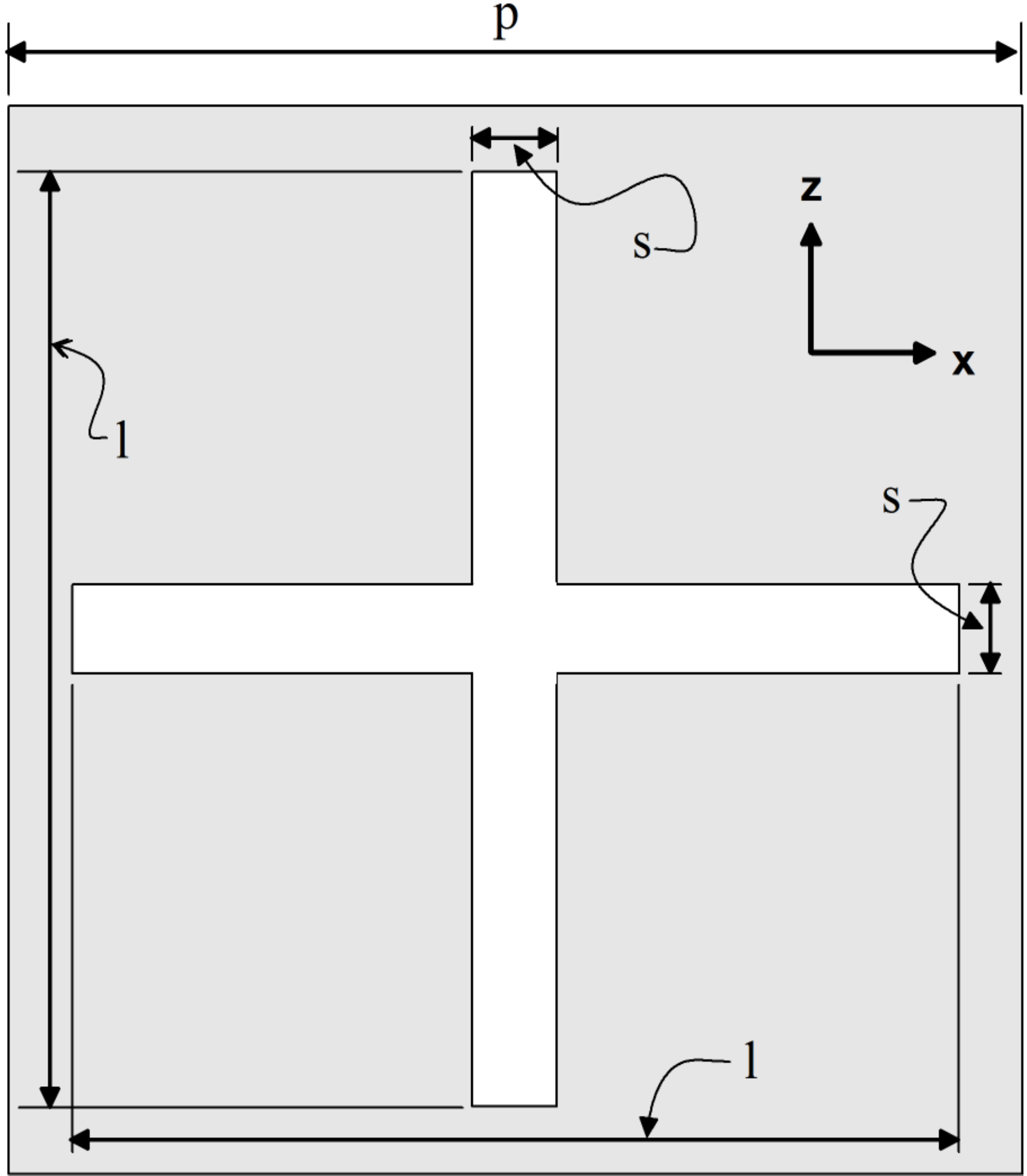}} \hspace{20mm}
\scalebox{.27}{\includegraphics{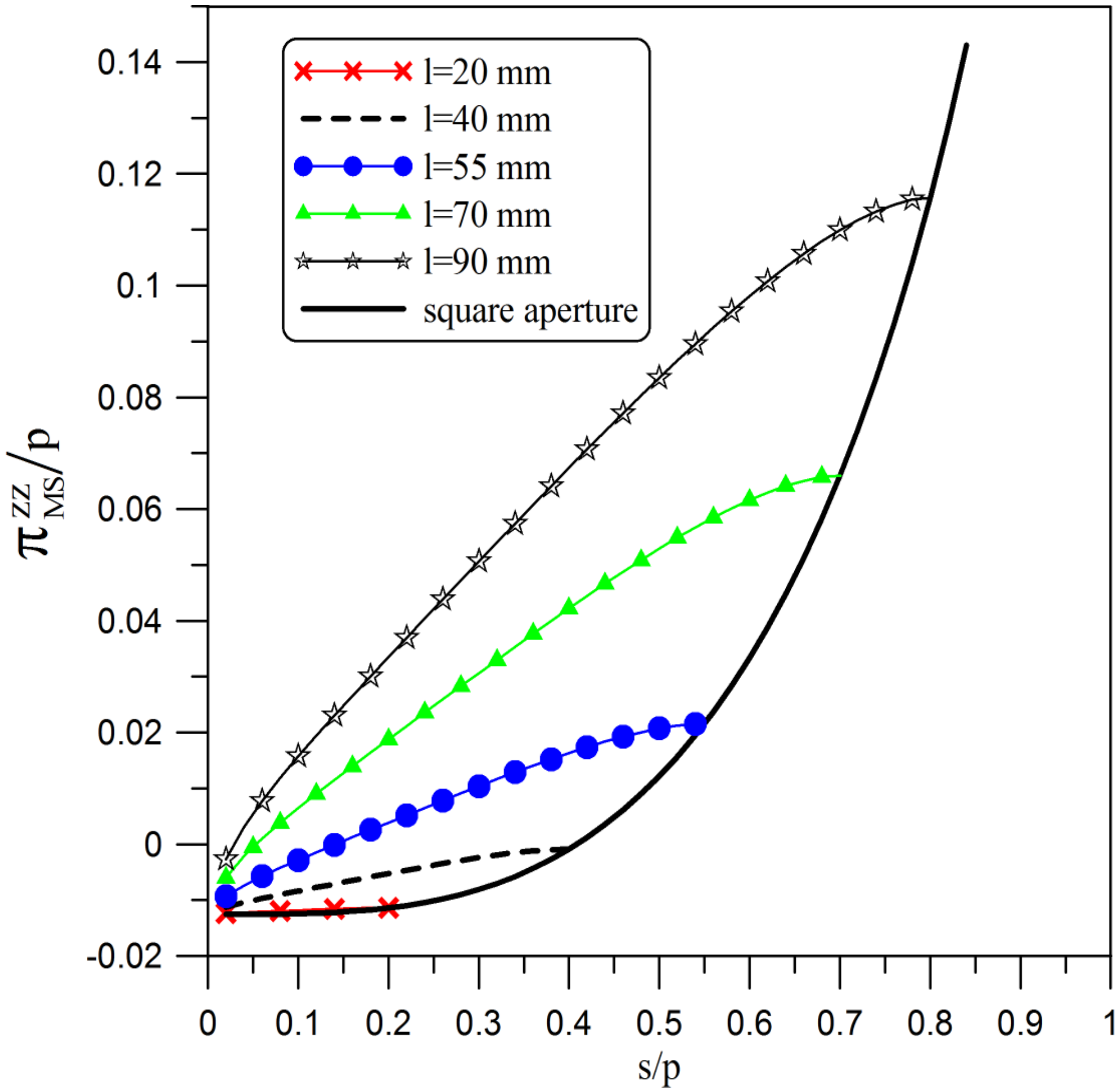}}\\
{\hspace{-1mm}\footnotesize{(a) \hspace{77mm} (b)}}\\
\vspace*{2mm}
\scalebox{.27}{\includegraphics{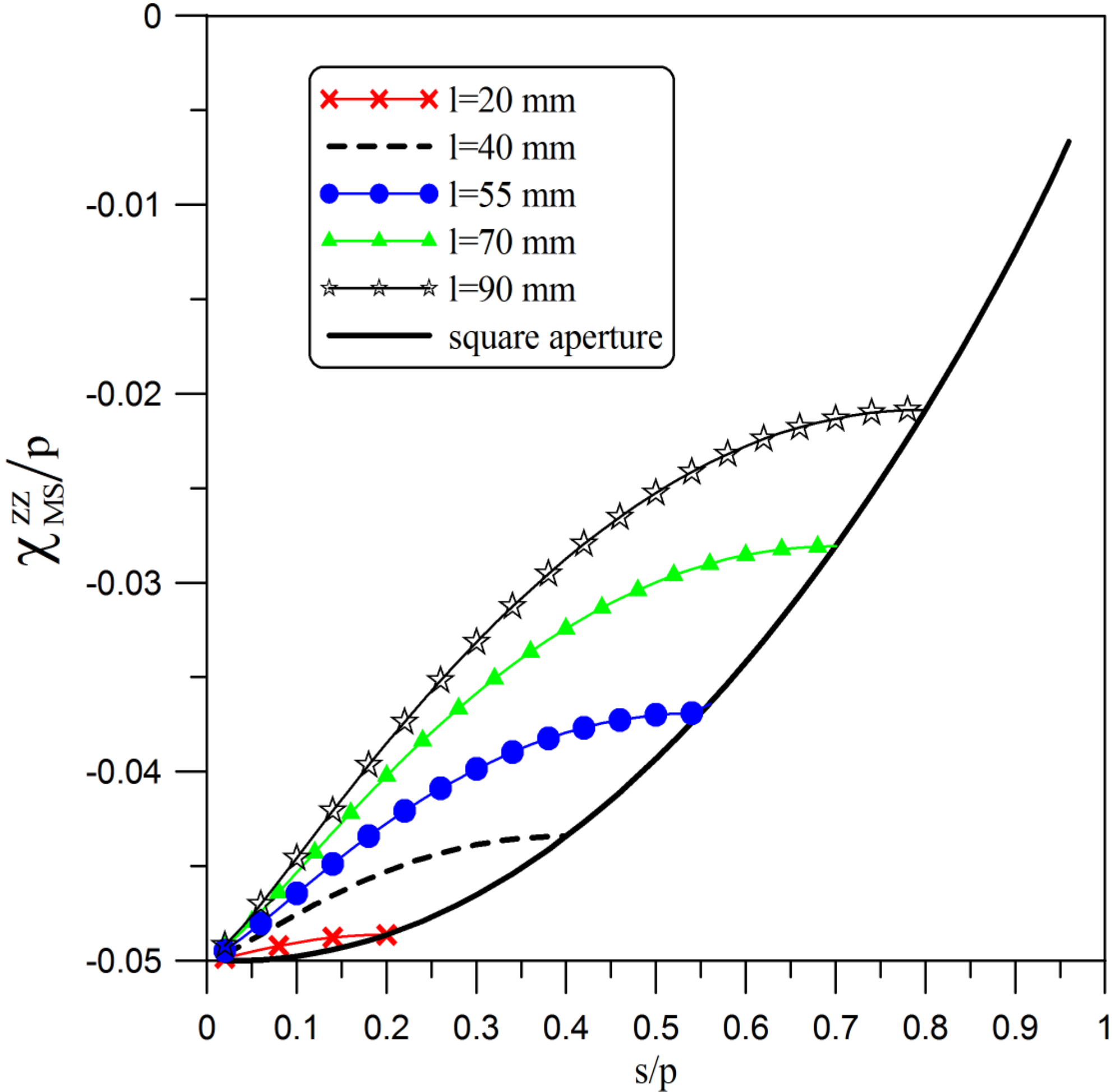}} \hspace{20mm}
\scalebox{.27}{\includegraphics{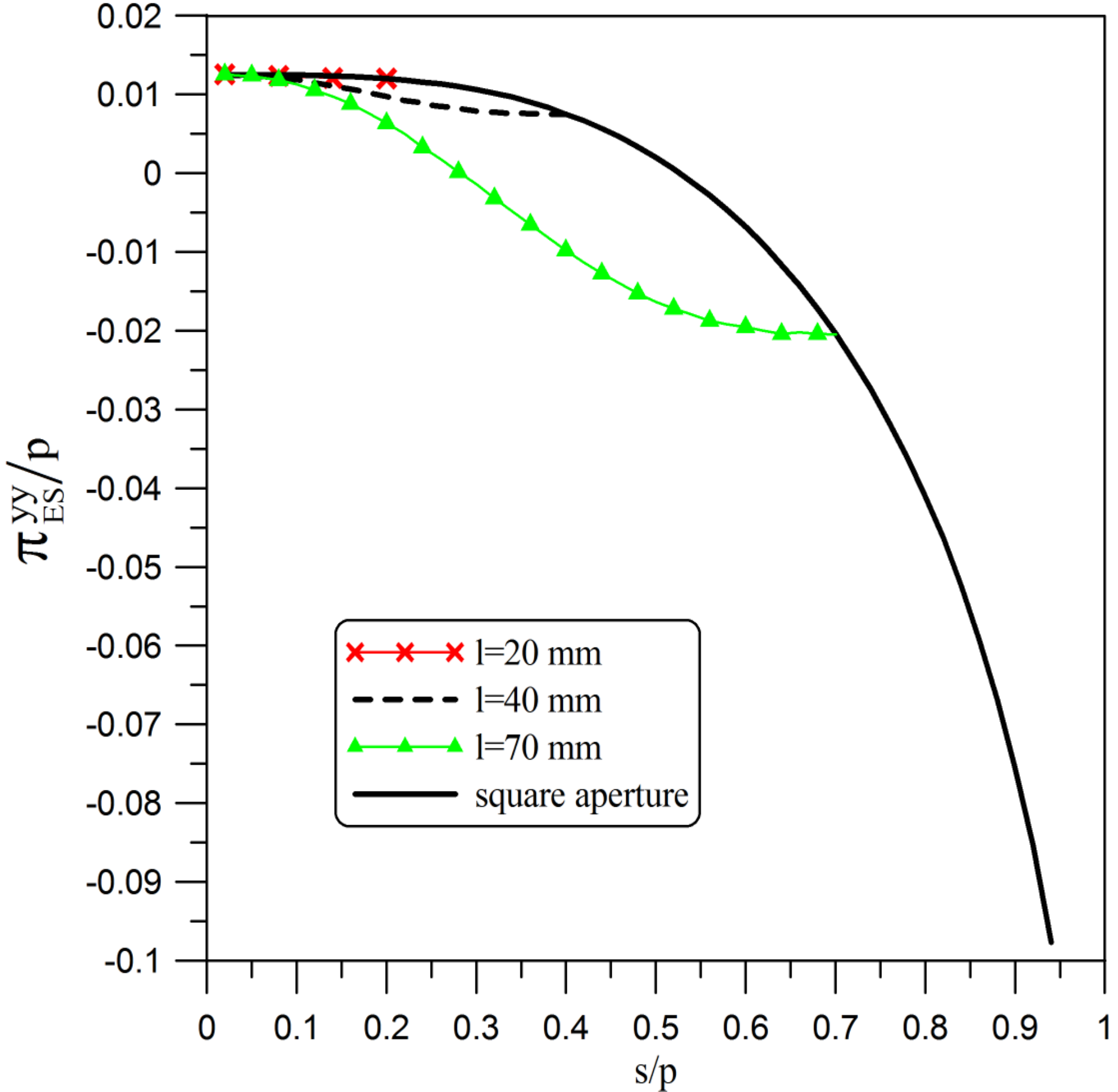}}\\
{\hspace{-1mm}\footnotesize{(c) \hspace{77mm} (d)}}\\
\vspace*{2mm}
\scalebox{.27}{\includegraphics{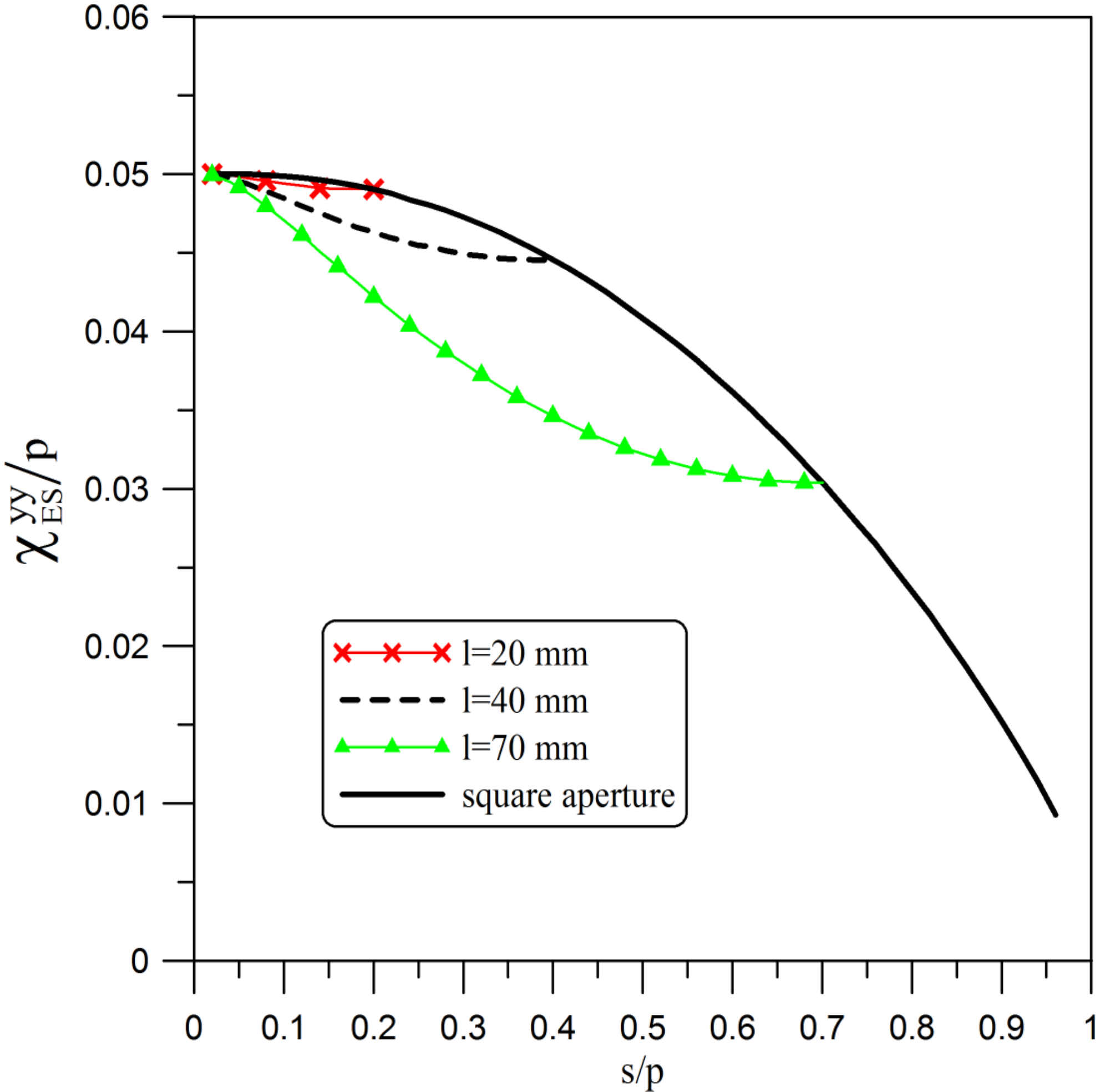}} 
\footnotesize{(e)}\\
\caption{Retrieved surface parameters for an array of cross-shaped apertures as a function of $s/l$ for a thickness $h=5$~mm and $p=100$~mm: (a) periodic cell of the slot, (b) $\pi_{MS}^{zz}$, (c) $\chi_{MS}^{zz}$, (d) $\pi_{ES}^{yy}$, and (e) $\chi_{ES}^{yy}$.}
\label{cross}
\end{figure*}

\begin{figure*}
\centering
\scalebox{.27}{\includegraphics{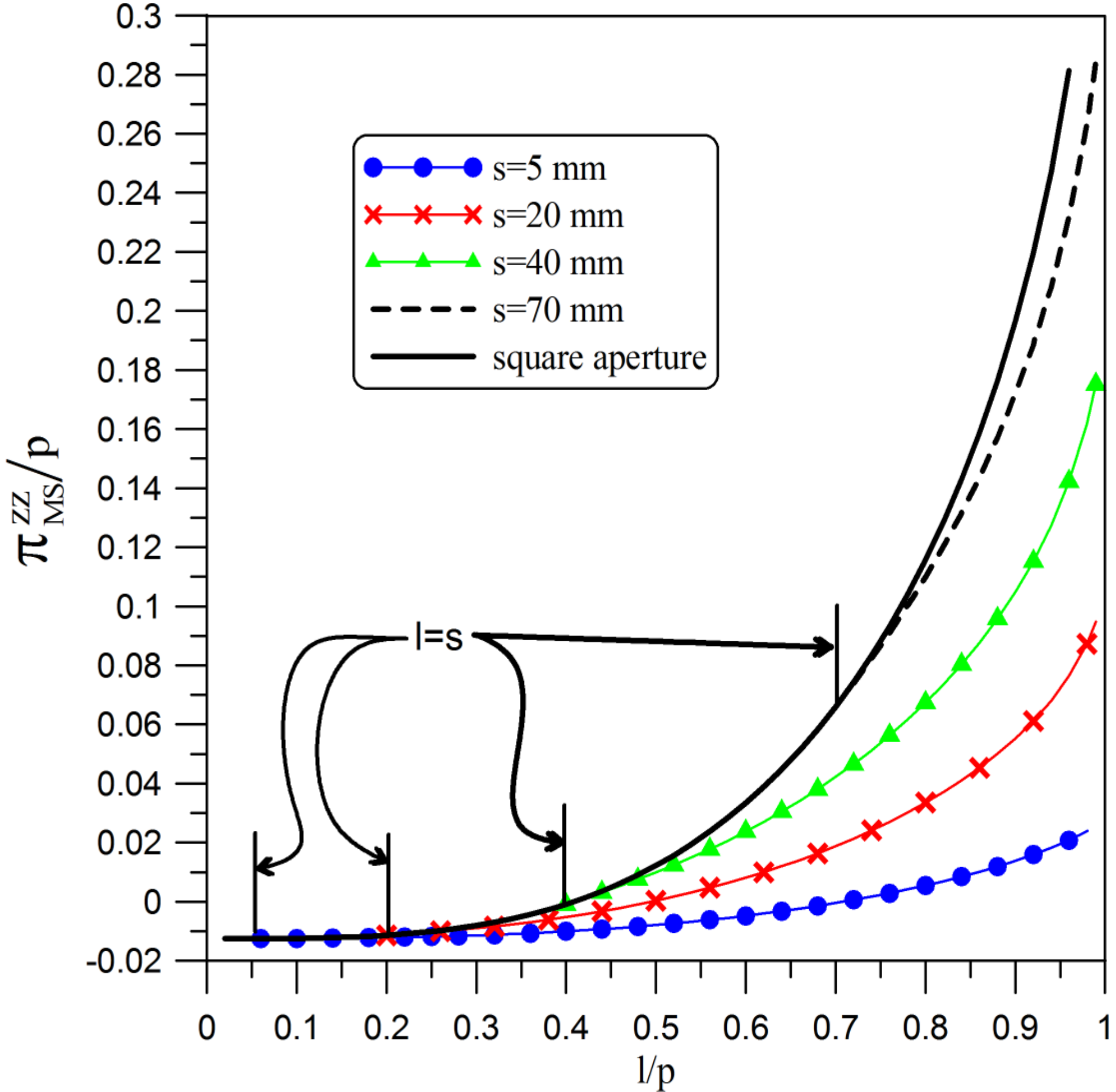}} \hspace{20mm}
\scalebox{.27}{\includegraphics{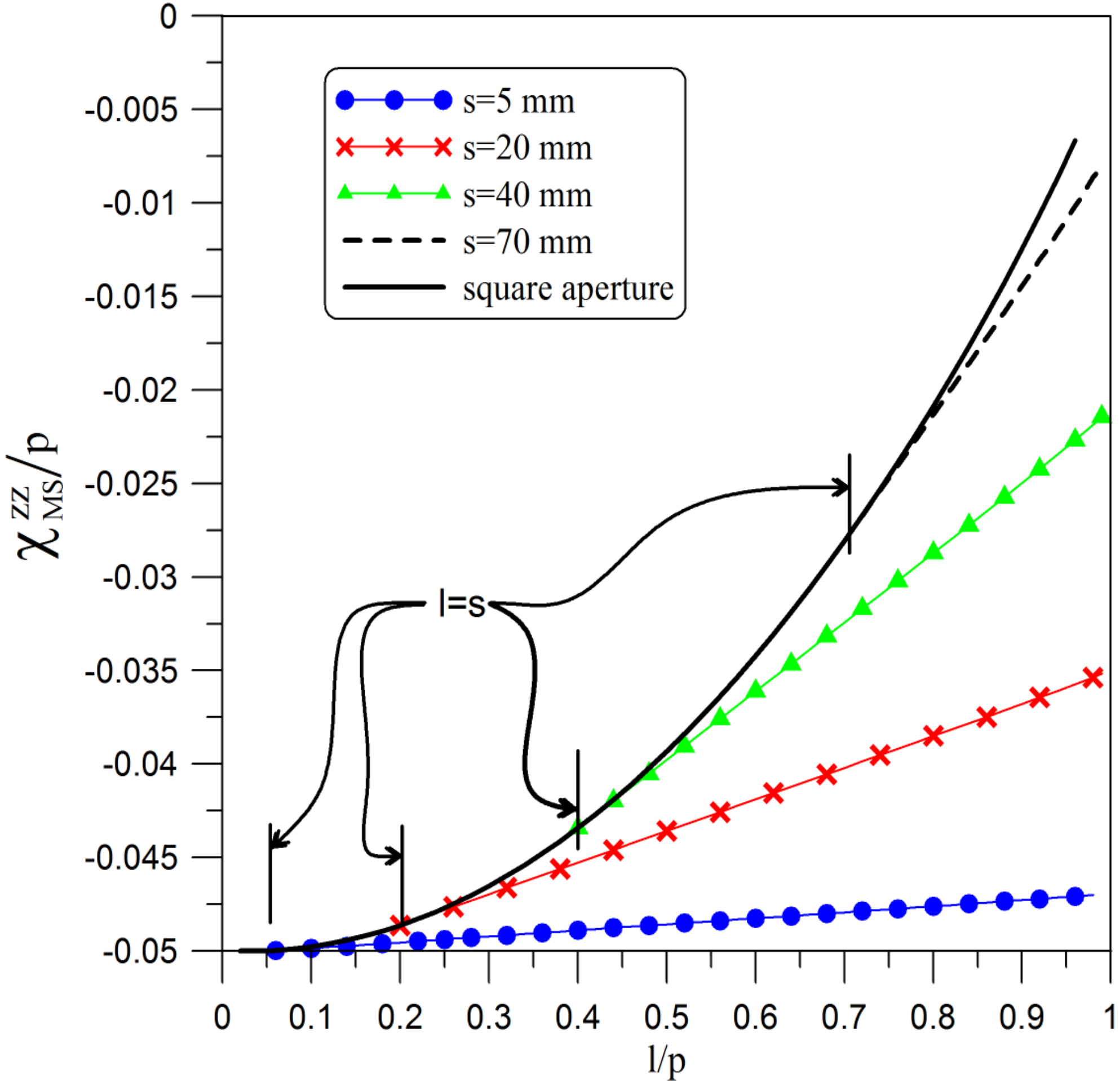}}\\
{\hspace{-1mm}\footnotesize{(a) \hspace{77mm} (b)}}\\
\vspace*{2mm}
\scalebox{.27}{\includegraphics{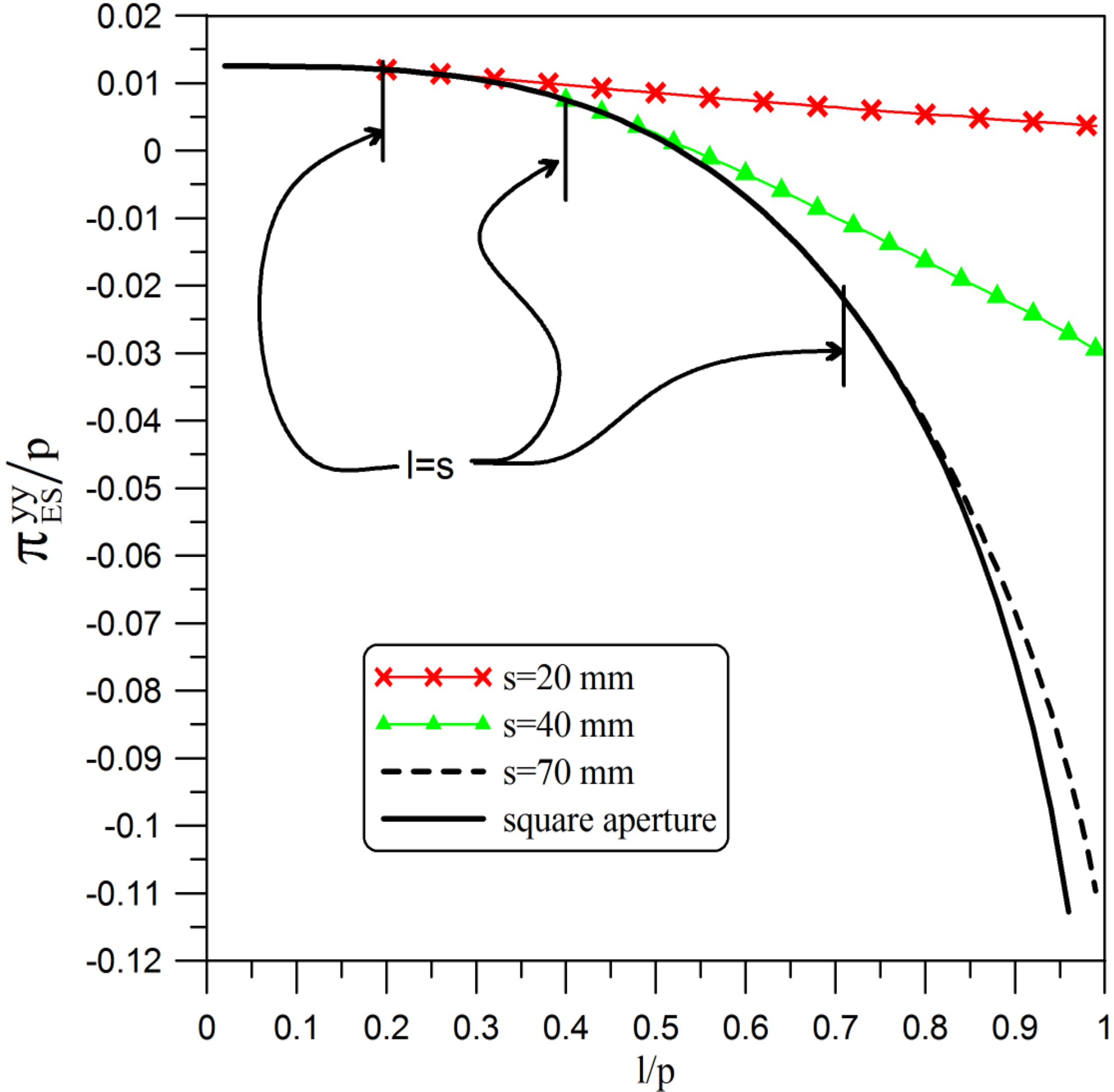}} \hspace{20mm}
\scalebox{.27}{\includegraphics{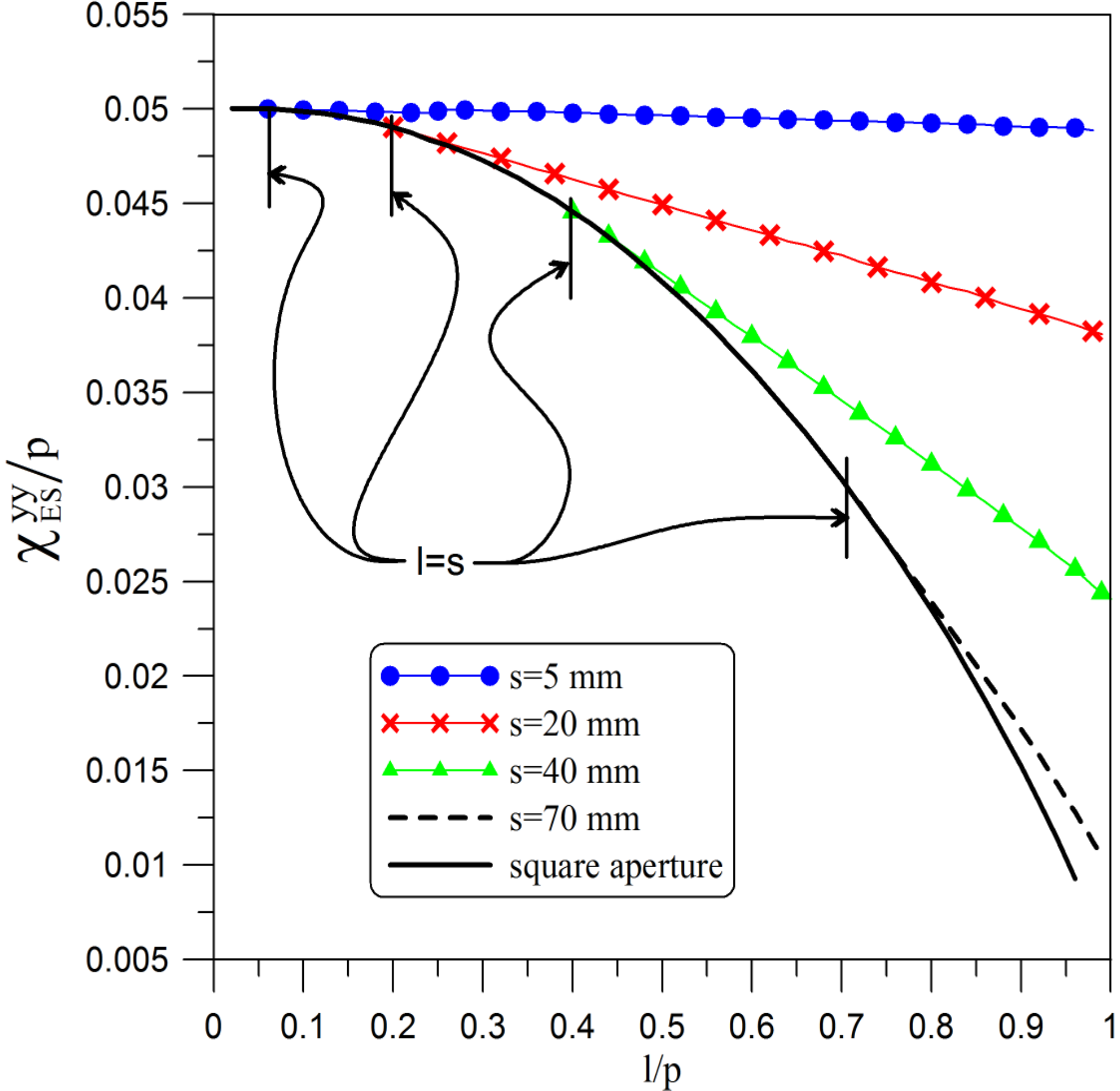}}\\
{\hspace{-1mm}\footnotesize{(c) \hspace{77mm} (d)}}\\
\caption{Retrieved surface parameters for an array of cross-shaped apertures as a function of $l/p$ for a thickness $h=5$~mm and $p=100$~mm:  (a) $\pi_{MS}^{zz}$, (c) $\chi_{MS}^{zz}$, (c) $\pi_{ES}^{yy}$, and (d) $\chi_{ES}^{yy}$}
\label{cross1}
\end{figure*}

\subsection{Metascreen Composed of an Array of Slot Apertures}
\label{slotsSec}

Next, we consider a metascreen composed of the slot apertures shown in Fig.~\ref{slot1}.  Here we show results for the surface parameters for various values of the slot length ($l$) and the slot width ($s$). In all these cases the length of the slot ($l$) is along the $x$-axis and the thickness of the apertures is 0.1~mm. In the limit as $l/p\rightarrow 1$ this structure reduces to an of array metallic strips, which will allow us to make comparisons to analytical values for $\pi_{MS}^{xx}$ as both $l/p\rightarrow 1$ and as $h\rightarrow 0$. The HFSS calculated reflection and transmission coefficients for this structure where used in eqs.~(\ref{pimzz})-(\ref{chieyy}) to obtained the surface parameters. Fig.~\ref{slot1}(b)-\ref{slot1}(g) shows results for the surface parameters as a function $l$ for different $s$. Note that $\pi_{MS}^{xx}\neq\pi_{MS}^{zz}$ and $\chi_{MS}^{xx}\neq\chi_{MS}^{zz}$ for this structure.
As $l/p\rightarrow 1$ this structure reduces to the flat metal strip grating analyzed in \cite{vain}.  In \cite{vain} Weinstein used five surface parameters in his boundary conditions and labeled them
as $l_0$, $l_1$, $l_2$, $l_3$, and $l_4$ and obtained approximate analytical expressions for them when the strip is infinity thin. Weinstein's parameter $l_0/2$ (see eq.~(2.96) in \cite{vain}) is equivalent to our $\pi_{MS}^{zz}$ when $l/p\rightarrow 1$ in our slot case (or more precisely $\pi_{MS}^{zz}\rightarrow l_0/2$ as $l/p\rightarrow 1$ and $h\rightarrow 0$).  In Fig.~\ref{slot1}(b) we also show $l_0/2$, and we see that $\pi_{MS}^{zz}$ approaches $l_0/2$ as $l/p\rightarrow 1$.   Also, for a thin strip grating $\chi_{MS}^{zz}\rightarrow l_2$ $\chi_{MS}^{xx}\rightarrow l_3$, and $\chi_{ES}^{yy}\rightarrow l_4$ as $l/p\rightarrow 1$ and $h\rightarrow 0$, Weinstein show $l_2=l_3=l_4\equiv 0$ when $h=0$. From Fig.~\ref{slot1}, we also see the the surface parameters $\chi_{MS}^{xx}$, $\chi_{MS}^{xx}$, and $\chi_{ES}^{yy}$ are very small and all approach zero as $l/p\rightarrow 1$ and $h\rightarrow 0$ (i.e., the thin slot array approaches a strip grating).


\begin{figure*}
\centering
\scalebox{.2}{\includegraphics{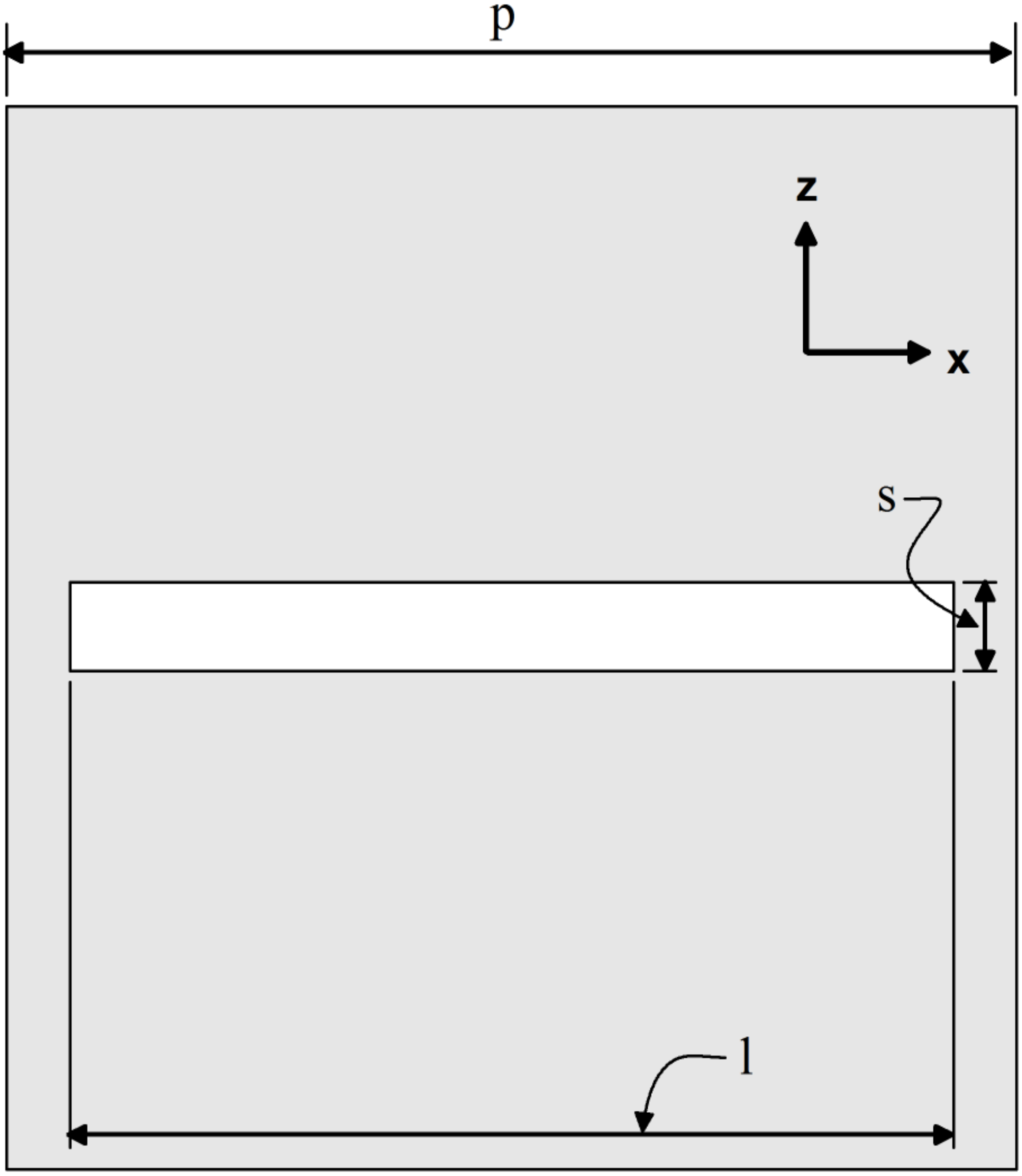}}\\
{\hspace{-1mm}\footnotesize{(a)}}\\
\scalebox{.27}{\includegraphics{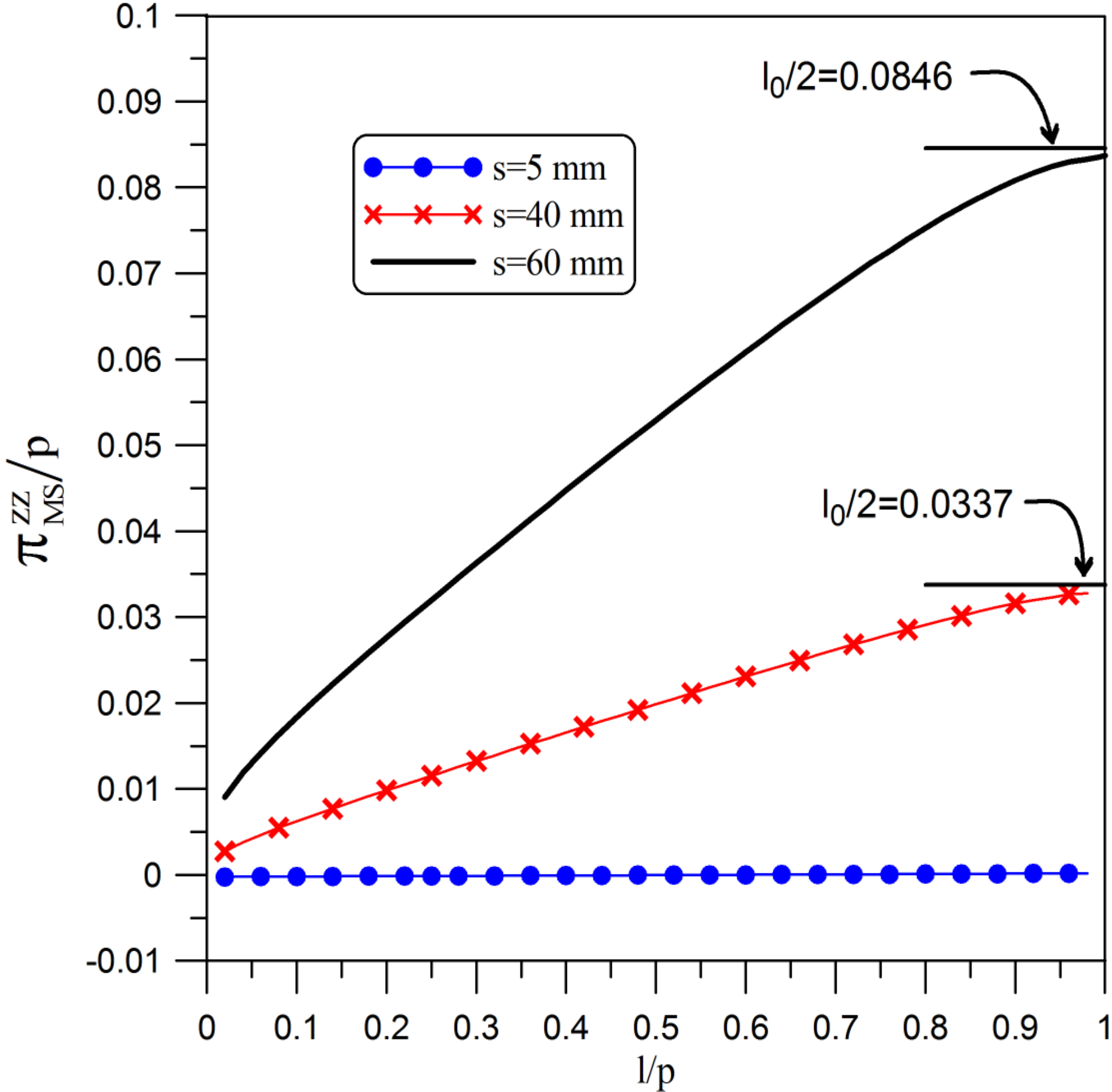}} \hspace{20mm}
\scalebox{.27}{\includegraphics{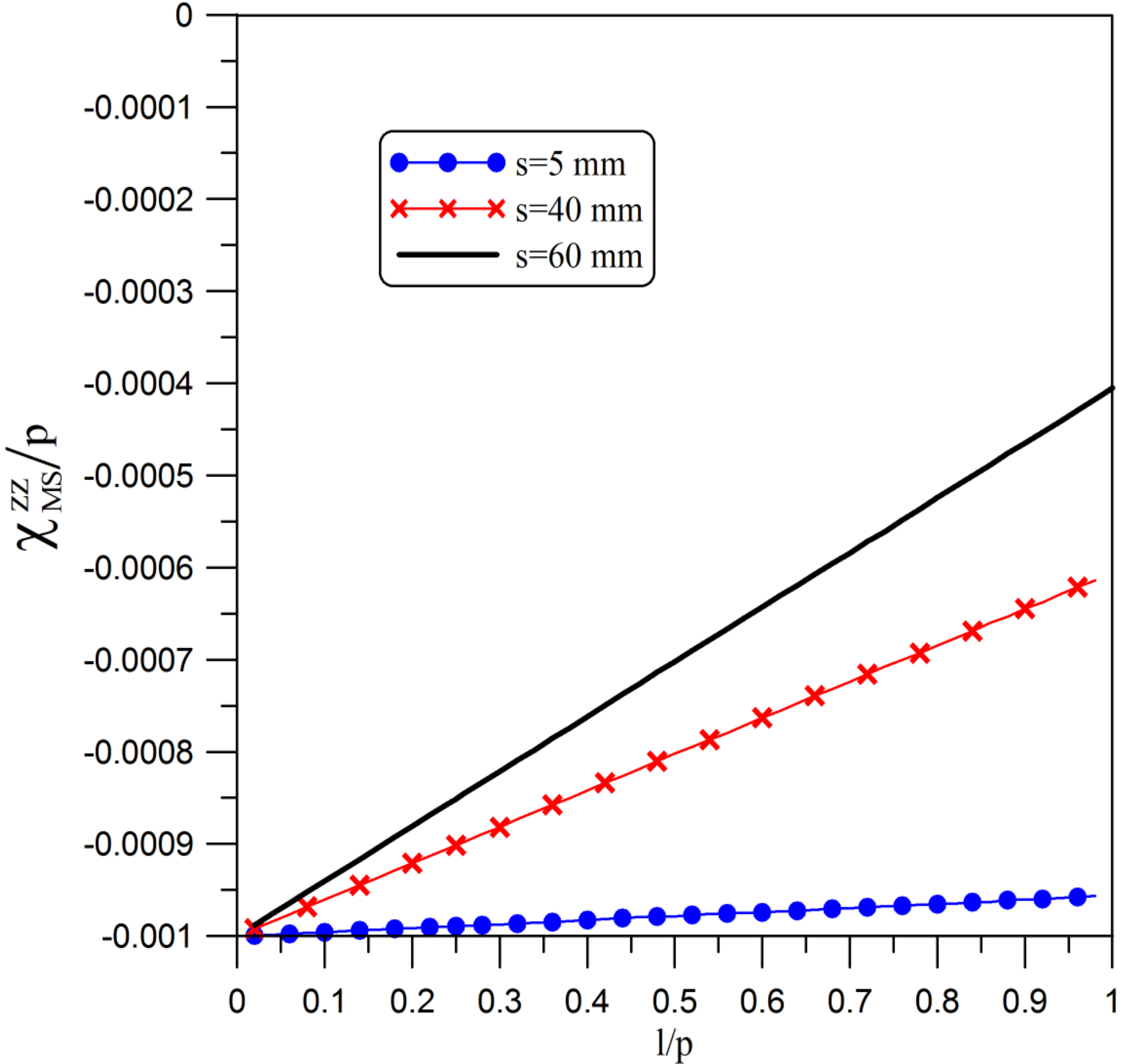}}\\
{\hspace{-1mm}\footnotesize{(b) \hspace{77mm} (c)}}\\
\vspace*{2mm}
\scalebox{.27}{\includegraphics{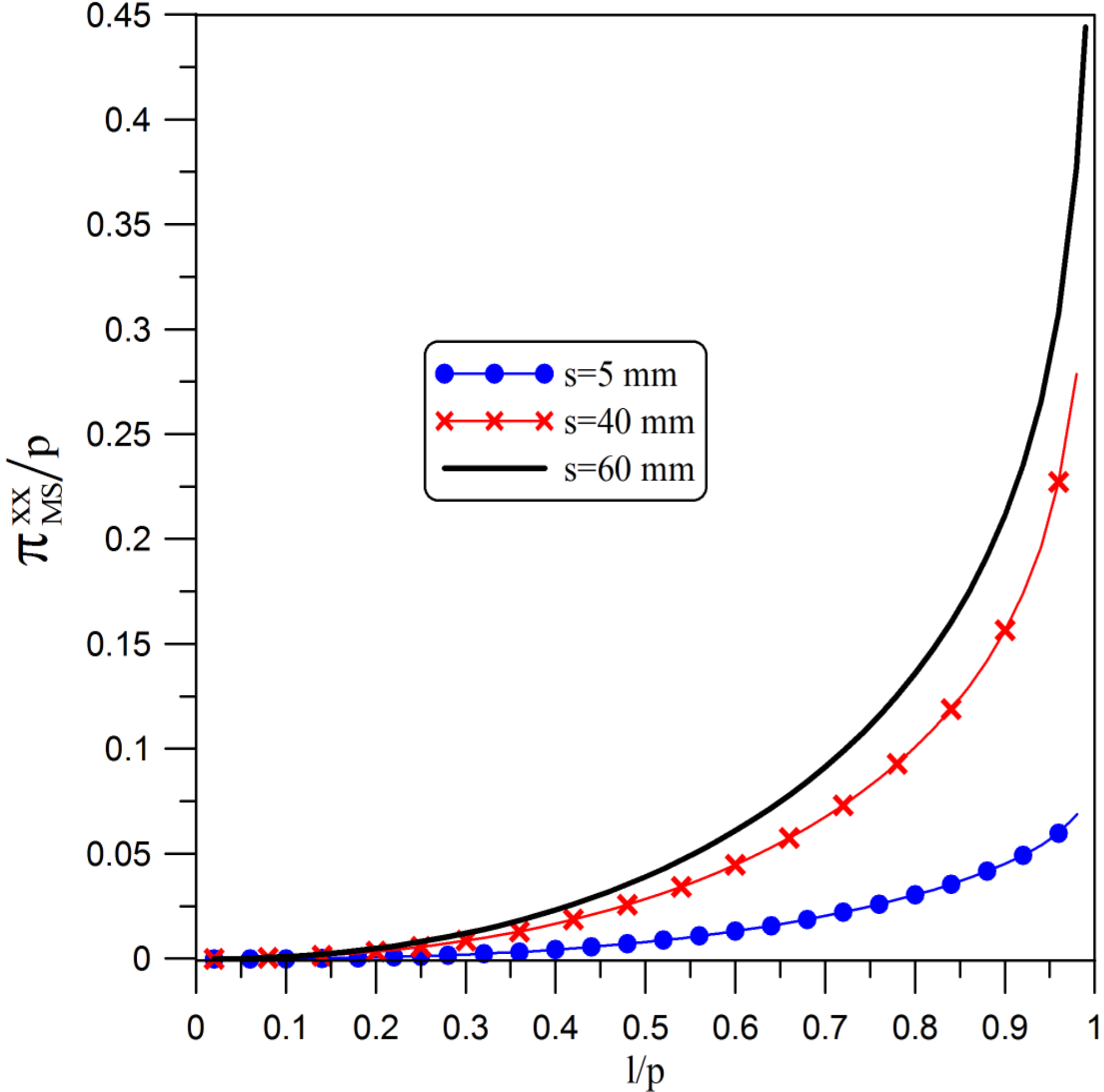}} \hspace{20mm}
\scalebox{.27}{\includegraphics{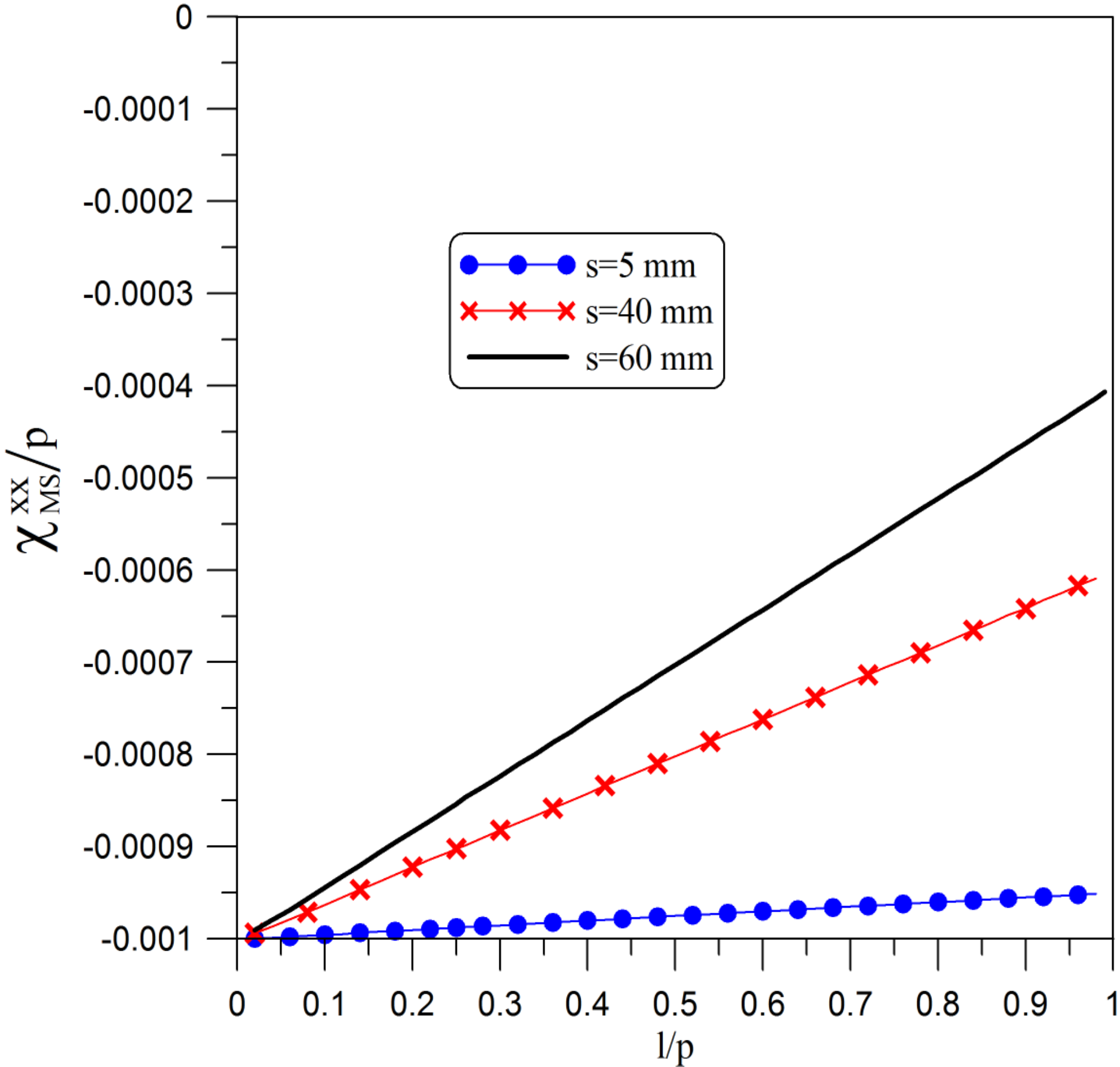}}\\
{\hspace{-1mm}\footnotesize{(d) \hspace{77mm} (e)}}\\
\vspace*{2mm}
\scalebox{.27}{\includegraphics{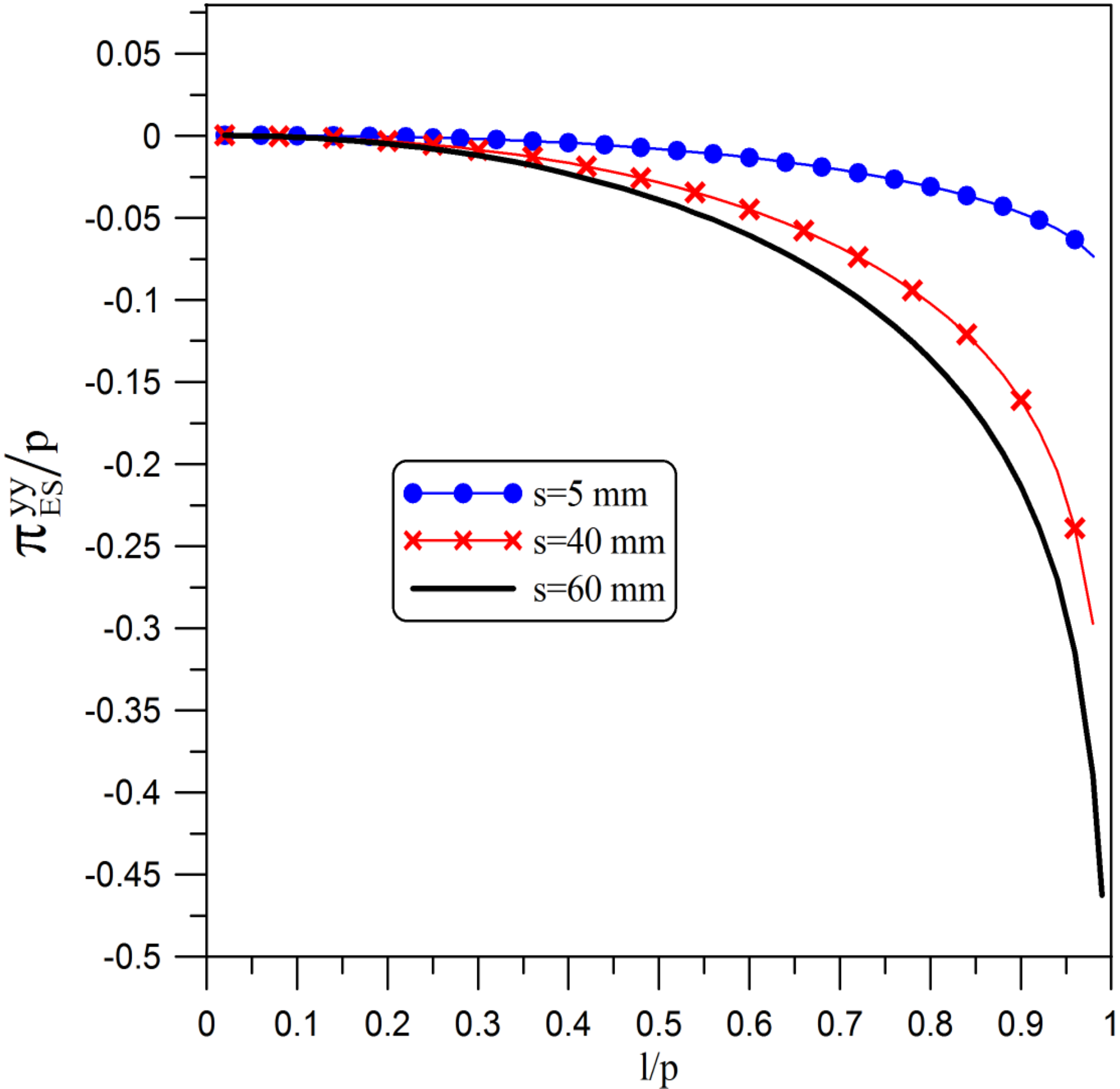}} \hspace{20mm}
\scalebox{.27}{\includegraphics{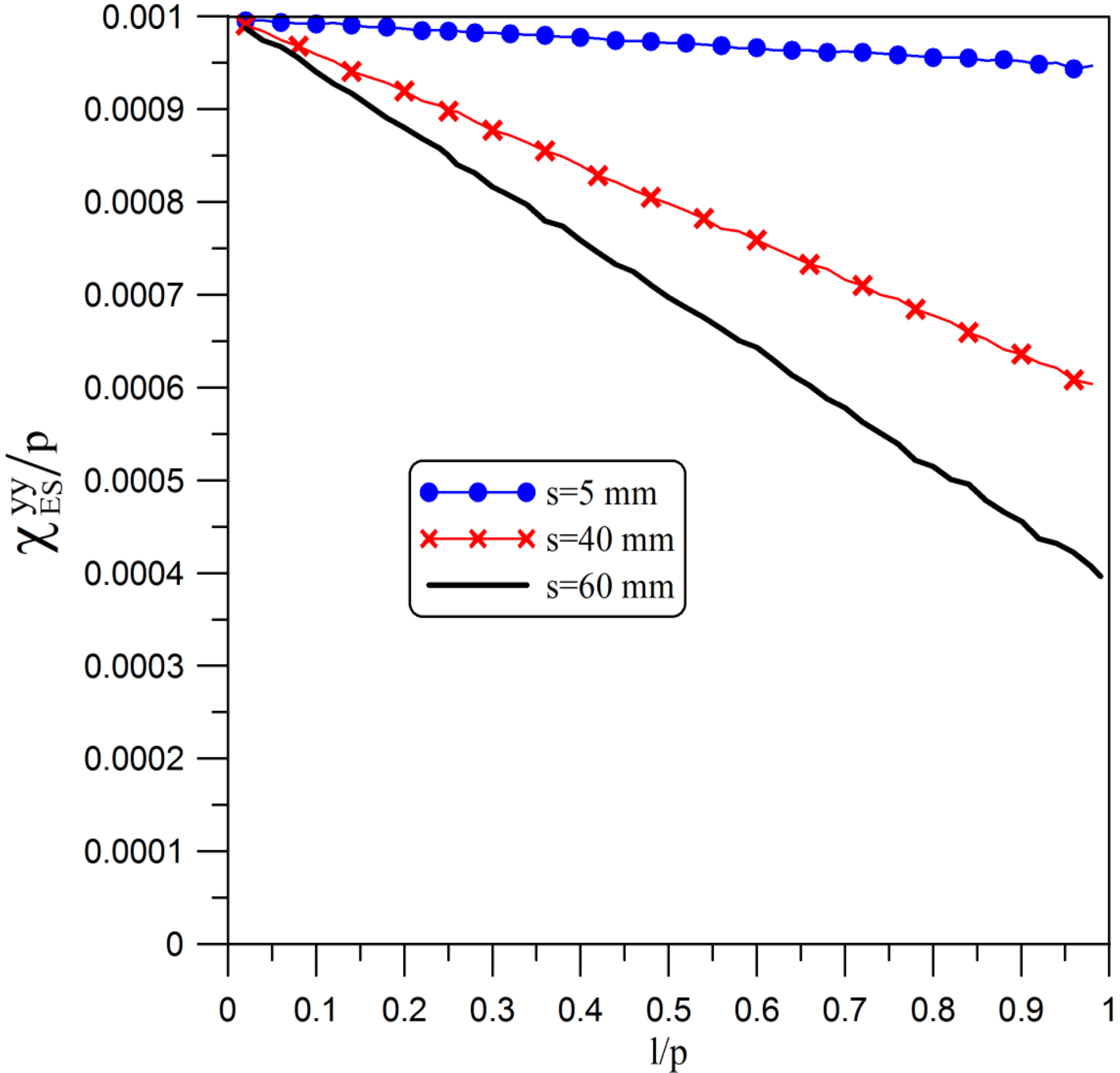}}\\
{\hspace{-1mm}\footnotesize{(f) \hspace{77mm} (g)}}\\
\caption{Retrieved surface parameters for an array of slots as a function of $l/p$ for a thickness $h=0.1$~mm:  (a) period cell, (b) $\pi_{MS}^{zz}$, (c) $\chi_{MS}^{zz}$, (d) $\pi_{MS}^{xx}$, (e) $\chi_{MS}^{xx}$, (f) $\pi_{ES}^{yy}$, and (g) $\chi_{ES}^{yy}$}
\label{slot1}
\end{figure*}

\subsection{Metascreen Composed of an Array of Square Apertures Filled with a High-contrast Dielectric}

In the last example, we consider a metascreen composed of an array of square apertures filled with a dielectric with $\epsilon_r=108.2$ and a loss tangent of $\tan\delta=4.9\times 10^{-5}$ (these materials properties represent a commercially available material).  This example illustrates three important points. First we show that the retrieval approach can also be used on a metascreen composed of lossy materials such that $\chi_{(ES,MS)}$ and $\pi_{(ES,MS)}$ have both ``real'' and ``imaginary'' parts. Secondly,  we illustrate that the retrieved surface parameters can be used to determine the transmission (or reflection) coefficient for any arbitrary angle of incidence.  Thirdly, we show that by filling the apertures with a high-contrast magneto-dielectric material (material with high-contrast electric and/or magnetic properties), it is possible to obtain interesting resonances at frequencies where no resonances exist when the aperture is not filled. Such behavior is not unexpected (after all, the apertures are filled with what are essentially dielectric resonators). Nevertheless, the retrieval technique developed in the present paper can obtain values for the surface susceptibilities and porosities, which in turn can be used to predict the behavior of a resonant metascreen under more general conditions (e.g., off-axis reflection and transmission coefficients, as shown below). This can lead to the possibility of designing metascreens with unique transmission and reflection properties (e.g., narrow-band filters as well as other applications).

Fig.~\ref{cubict} shows the normal-incidence transmission coefficient as a function of frequency for an array of square apertures filled with a dielectric material ($\epsilon_r=108.2$ and $\tan\delta=4.9\times 10^{-5}$). These results were obtained with HFSS for $p=26$~mm, $h$=10~mm, and $l=10$~mm, see Fig.~~\ref{array}(b).  In Fig.~\ref{cubict}, we also show results for the case when no material is in the apertures. Comparing the two results in this figure, we see that the presence of the material filling the aperture caused two resonances (at 1.63~GHz and 2.33~GHz) which are not present when no material is used. We see that at 1.63~GHz and 2.33~GHz, the surface allows almost 100~$\%$ transmission (the losses cause this to be $<100$~$\%$) over two narrow frequency bands.  These two high-transmission regions are a result of internal resonances in the high-contrast material filling the apertures.

By retrieving the various surface parameters, we can get an indication as to which of the surface parameters give rise to the resonances in the transmission coefficient.
Using the HFSS results for $R_{TE}(0)$ and $T_{TE}(0)$ for this metascreen, we used eqs.~(\ref{pimzz}) and (\ref{chimzz}) to determine $\pi_{MS}^{xx}$ and $\chi_{MS}^{xx}$. These retrieved values are shown in Fig.~\ref{chimpim}. Shown here are the real and imaginary parts of the surface parameters (where the imaginary parts arise due to the loss tangent of the material). Note that in these figures we have only shown results zoomed in around the two resonant frequencies.  From the figure, we see that $\pi_{MS}^{xx}$ exhibits a resonance at 1.63~GHz, while  $\chi_{MS}^{xx}$ has a flat response (and is about two orders of magnitude smaller) in this same frequency range.  The resonance in $\pi_{MS}^{xx}$  is what gives rise to the enhanced transmission at 1.63~GHz. On the other hand,  $\chi_{MS}^{xx}$ exhibits a resonance at 2.33~GHz ($\pi_{MS}^{xx}$ does not and is about three orders of magnitude smaller) and  $\chi_{MS}^{xx}$ is what gives rise to the enhanced transmission at 2.33~GHz.

\begin{figure}
\centering
\scalebox{0.3}{\includegraphics*{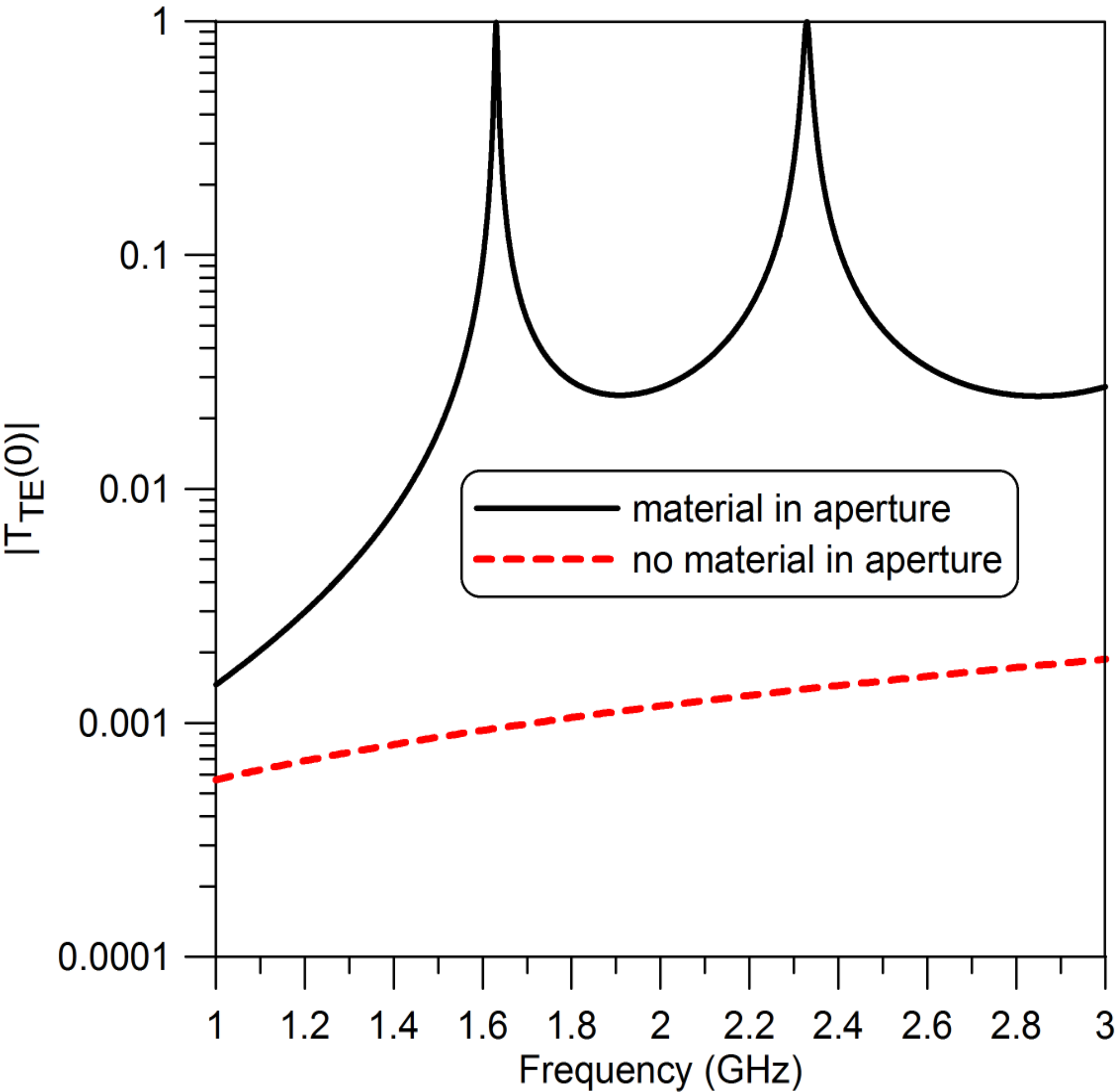}}
\caption{HFSS results for the transmission coefficient for an array of square apertures: $p=26$~mm, $h=10$~mm, and $l=10$~mm.}
\label{cubict}
\end{figure}

\begin{figure}
\centering
\scalebox{.32}{\includegraphics{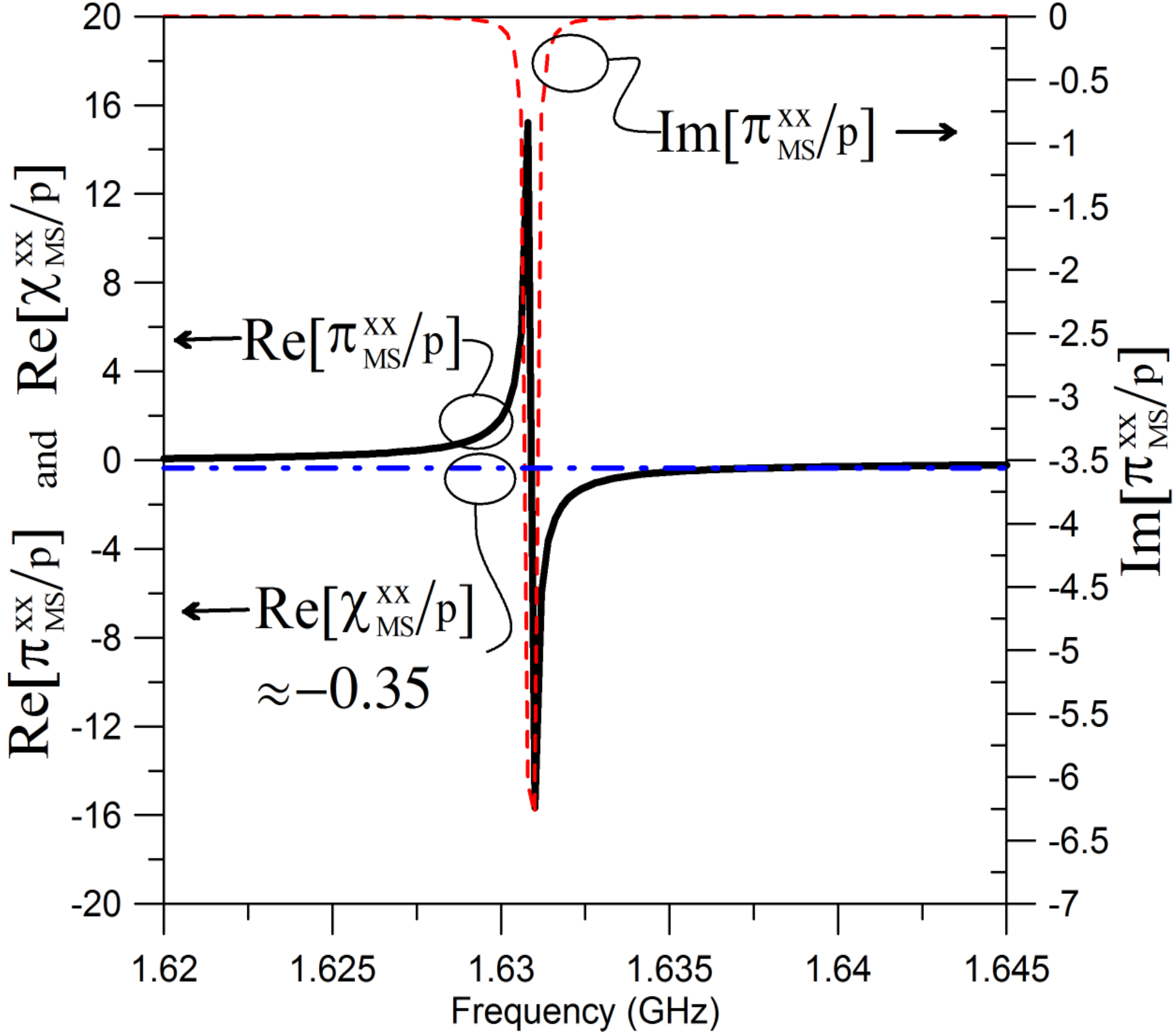}}\\ {\hspace{-1mm}\footnotesize{(a) }}\\
\vspace{6mm}
\scalebox{.32}{\includegraphics{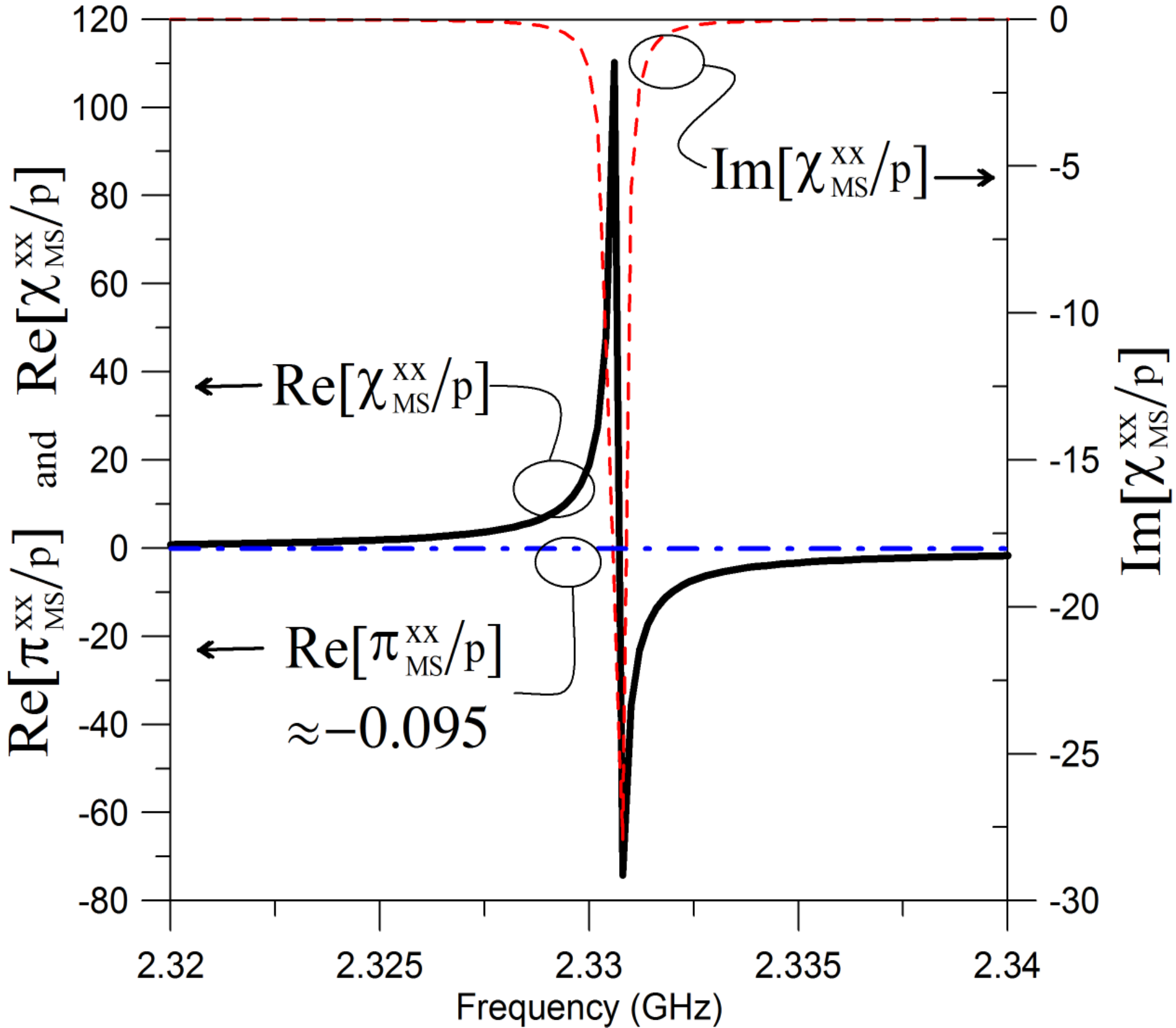}}\\ {\hspace{-1mm}\footnotesize{(b)}}\\
\caption{Retrieved surface parameters ($\pi_{MS}^{xx}$ and $\chi_{MS}^{xx}$) for an array of square aperture filled with a high-contrast material ($\epsilon_r=108.2$ and $\tan\delta=4.9\times 10^{-5}$) for $p=26$~mm, $h=10$~mm, and $l=10$~mm:  (a) 1.63~GHz and (c) 2.33~GHz.}
\label{chimpim}
\end{figure}

Using these retrieved values for $\pi_{MS}^{xx}$ and $\chi_{MS}^{xx}$, along with eq.~(\ref{tet}), the transmission coefficient for various angles of incidence can be determined (recall that only two surface parameters are needed to determine the transmission coefficient for a TE wave). The values for $T_{TE}(\theta)$ calculated from (\ref{tet}) and the results given in Fig.~\ref{chimpim} are shown in Fig.~\ref{figtet} for 30$^0$, 45$^0$, and $60^0$. For a comparison, also in this figure, we show numerical results obtained from HFSS.  These comparisons show that the transmission coefficient obtained from surface parameters retrieved from normal incidence data are indistinguishable from the HFSS results, even for angles as high as 60$^{\circ}$.

\begin{figure}
\centering
\scalebox{0.3}{\includegraphics*{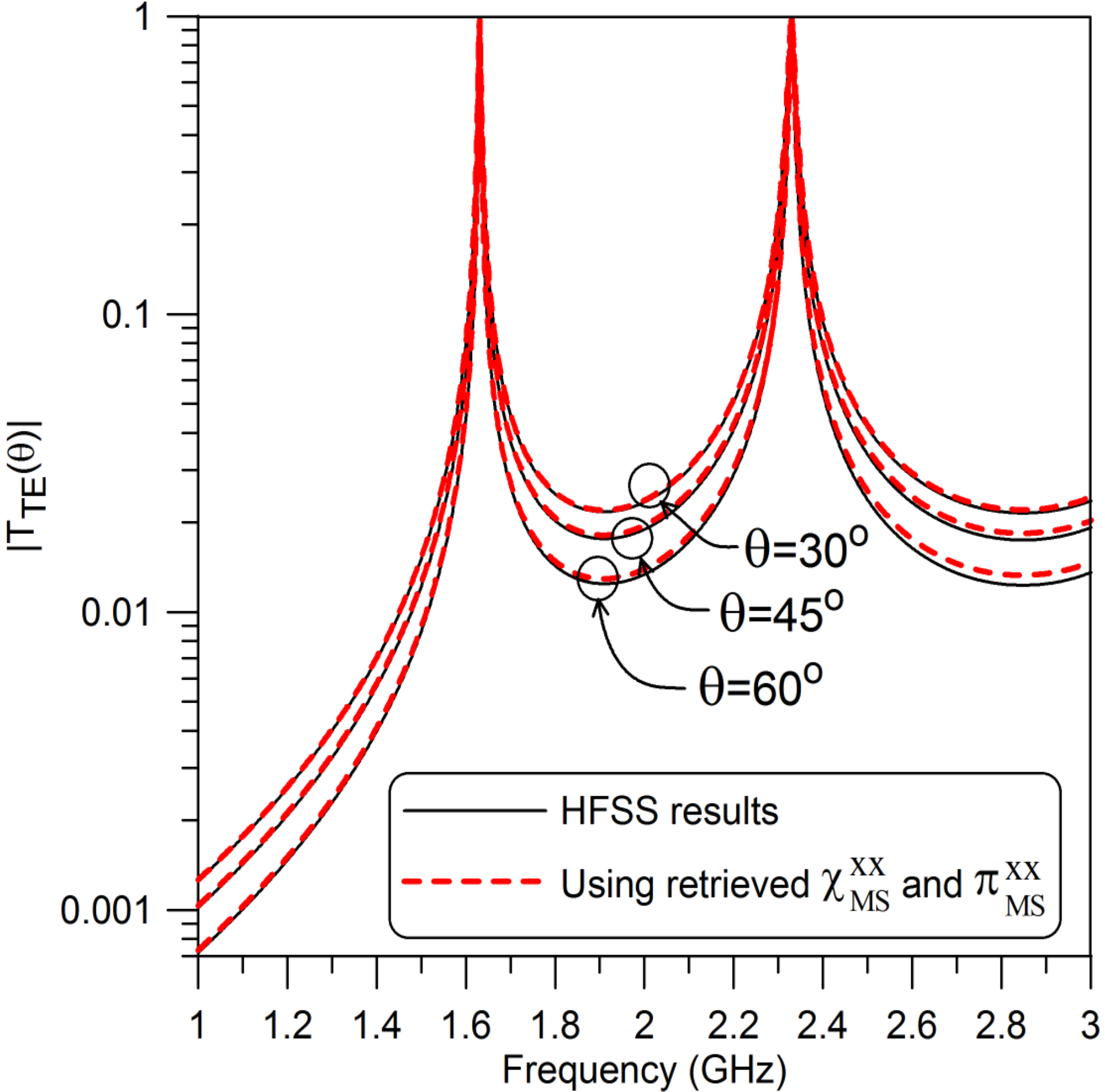}}
\caption{Comparison of $T_{TE}(\theta)$ for an array of square aperture filled with a high-contrast ($\epsilon_r=108.2$ and $\tan\delta=4.9\times 10^{-5}$) for $p=26$~mm, $h=10$~mm, and $l=10$~mm.}
\label{figtet}
\end{figure}

As shown in (\ref{tmt}), $T_{TM}(\theta)$ for an arbitrary angle of incidence requires four surface parameters ($\pi_{MS}^{zz}$, $\chi_{MS}^{zz}$, $\pi_{ES}^{yy}$, and $\chi_{ES}^{yy}$).  Since this metascreen is symmetric, $\pi_{MS}^{zz}=\pi_{MS}^{xx}$ and
$\chi_{MS}^{zz}=\chi_{MS}^{xx}$ (which are given in Fig.~\ref{chimpim}). To obtain $\pi_{ES}^{yy}$ and $\chi_{ES}^{yy}$, we used (\ref{pieyy}) and (\ref{chieyy}) along with HFSS numerical values for  $T_{TM}(\theta)$ at $30^{\circ}$ (the values for $|T_{TM}(30^{\circ})|$ are shown in Fig.~\ref{figtmt}).
Although not shown here, $\pi_{ES}^{yy}$ also has a resonance at 1.63~GHz, and $\chi_{ES}^{yy}$ also has a resonance at 2.33~GHz. Using these four retrieved surface parameters, along with eq.~(\ref{tmt}), the transmission coefficient for various angles of incidence can be determined. The calculated values for $T_{TM}(\theta)$ from (\ref{tmt}) for 30$^{\circ}$, 45$^{\circ}$, and $60^{\circ}$ are shown in Fig.~\ref{figtmt}. For a comparison, we also show numerical results obtained from HFSS.  This comparison shows that the transmission coefficients obtained from the retrieved surface parameters are indistinguishable from the HFSS results, even for angles as high as 60$^{\circ}$. In fact, the ``{\it Fano}'' type resonances that occur around 2.4~GHz, 2.5~GHz, and 2.8~GHz are also captured using the retrieved surface parameters (Fig.~\ref{figzoom} shows results zoomed in near these resonances). The differences in these very narrow Fano resonant frequencies do not exceed $0.05~\%$.

\begin{figure}
\centering
\scalebox{0.3}{\includegraphics*{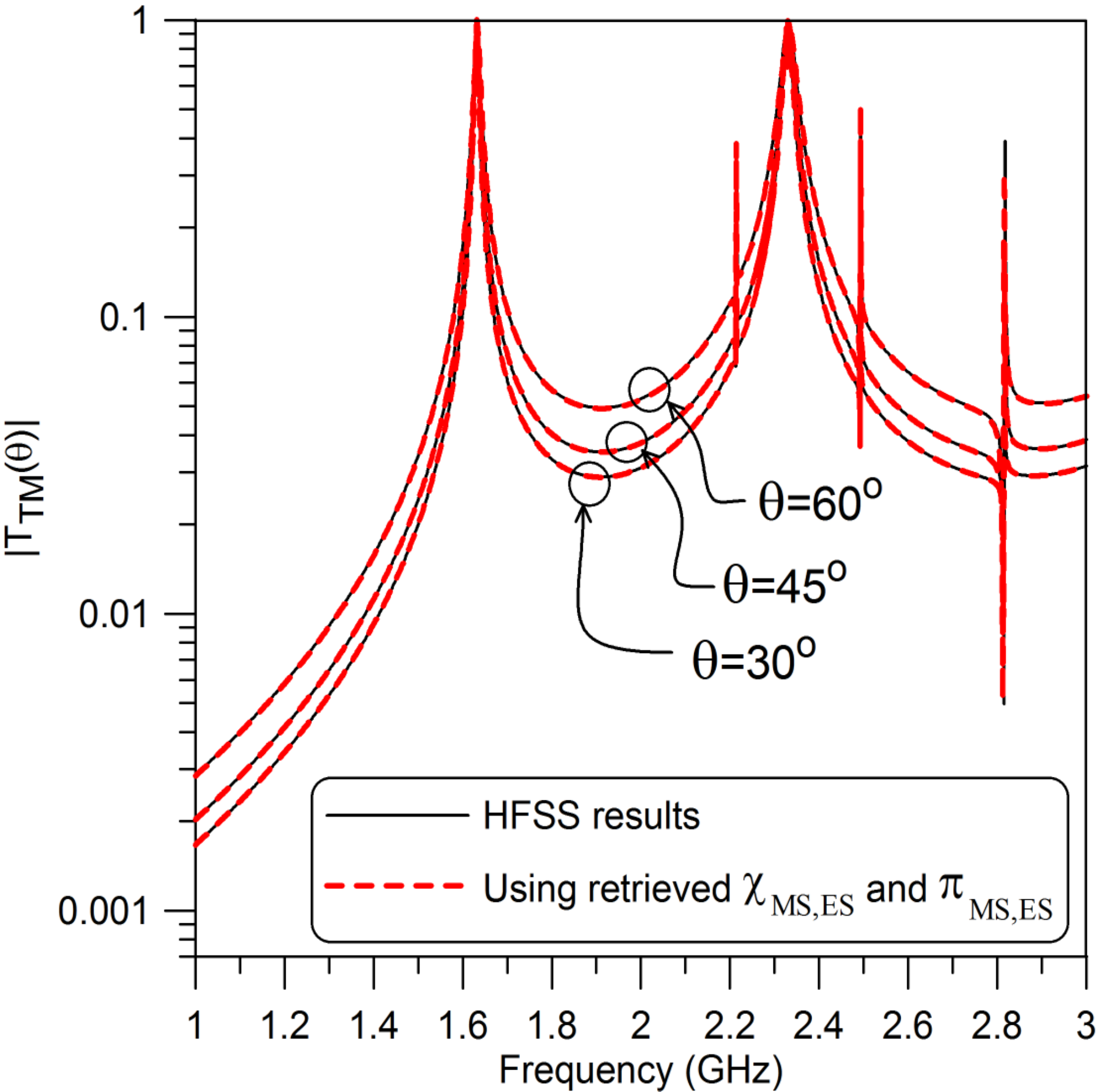}}
\caption{Comparison of $T_{TM}(\theta)$ for an array of square aperture filled with a high-contrast ($\epsilon_r=108.2$ and $\tan\delta=4.9\times 10^{-5}$) for $p=26$~mm, $h=10$~mm, and $l=10$~mm.}
\label{figtmt}
\end{figure}

\begin{figure}
\centering
\scalebox{0.34}{\includegraphics*{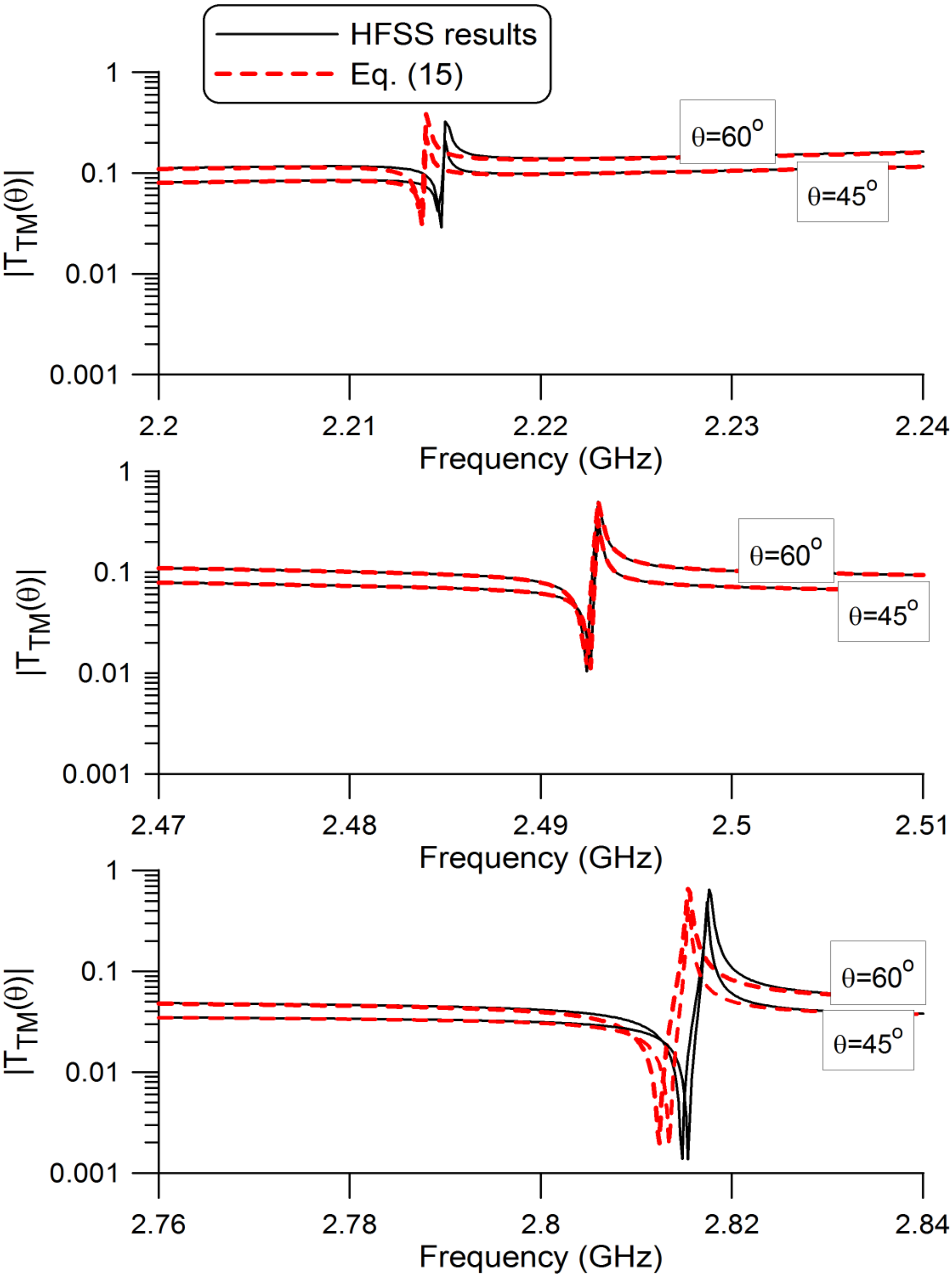}}
\caption{Zoomed-in region of {\it Fano} type resonances in $T_{TM}(\theta)$ at $45^{\circ}$ and $60^{\circ}$.}
\label{figzoom}
\end{figure}


\subsection{Limit as the Surface-area of the Apertures Diminish}

It is interesting to observe the surface parameters' behavior
when the surface area of the apertures go to zero, or when the fill-factor (either $l/p$ or $a/p$) approaches zero. Referring to the results in the various Figures, we see that as $l/p\rightarrow 0$ (or as $a/p\rightarrow 0$),
\begin{equation}
\begin{array}{c}
\pi_{ES}^{Ayy}=\pi_{ES}^{Byy} \equiv 2 \pi_{ES}^{yy}\rightarrow h/2\\
\pi_{MS}^{Axx}=\pi_{MS}^{Bxx} \equiv \,\,\,\, 2 \pi_{MS}^{xx}\rightarrow -h/2\\
\pi_{MS}^{Azz}=\pi_{MS}^{Bzz} \equiv \,\,\,\, 2 \pi_{MS}^{zz}\rightarrow -h/2\\
\end{array} \,\, ,
\label{symb22}
\end{equation}
and
\begin{equation}
\begin{array}{c}
\chi_{ES}^{Ayy}=\chi_{ES}^{Byy} \equiv \frac{1}{2} \chi_{ES}^{yy}\rightarrow h/2\\
\chi_{MS}^{Axx}=\chi_{MS}^{Bxx} \equiv \frac{1}{2} \chi_{MS}^{xx}\rightarrow -h/2\\
\chi_{MS}^{Azz}=\chi_{MS}^{Bzz} \equiv \frac{1}{2} \chi_{MS}^{zz}\rightarrow -h/2\\
\end{array} \,\, .
\label{symb33}
\end{equation}
These zero fill-factor limits are also obtained by observing how the surface parameters are defined in \cite{metascreen} (see eqs.~(73), (78) and (80) therein). The integrals involving the static field given in \cite{metascreen} are zero for $l/p=0$ (or for $a/p=0$), and hence the surface parameters reduce to $\pm h/2$.  The physical reason for this is because of the choice of the reference plane location. This is made clear by looking at the reflection and transmission coefficients under the zero fill-factor condition. Using (\ref{symb22}) and (\ref{symb33}) it can be shown that in this limit, eqs.~(\ref{ter}) and (\ref{tet}) reduce to:
\begin{equation}
\begin{array}{rcl}
R_{TE}(\theta)=e^{-j\phi}&;&
T_{TE}(\theta)=0\\
\end{array}
\label{tertetzero}
\end{equation}
where
\begin{equation}
\small
\phi=\tan^{-1}\left(\frac{h\,k_0\cos\theta}{\left(\frac{h\,k_0\,\cos\theta}{2}\right)^2-1}\right)\sim h\,k_0\cos\theta+O\left( (h\,k_0)^3\right)\,\, .
\label{phiangle}
\end{equation}
From Fig.~\ref{fig2} and  \cite{metascreen}, the reference plane is at  $y=0$  (i.~e., half way between the top and bottom surfaces of the conducting plane). This reference plane represents the location where the GSTCs are applied. When no apertures are present in the plane, this would correspond to a phase shift of $h\,k_0\cos\theta$ for a reflected field relative to a position on the top side of the plane. This would correspond to the phase shift of a wave propagated from the top surface of the metascreen to $y=0$ and back. If we had chosen a reference plane at some other location, the surface parameters for zero fill-factor would change accordingly. In any event, as $h\rightarrow 0$, this phase shift would also approach zero, and we would obtain the result for the reflection for a conducting plane as if a reference plane was on the surface (i.e., no phase shift).

\subsection{Limit as the Apertures Touch}

The other extreme limit of interest is when $l/p\rightarrow 1$ or $a/p\rightarrow 1/2$. For the square aperture, $l/p\rightarrow 1$ corresponds to the square apertures touching and as a result the screen disappearing (i.~e., no conducting screen is present). From Figs.~\ref{pimchim2}, and \ref{piechie2}, we see that for this limit, $\chi_{ES}$ and $\chi_{MS}$ approach zero, and $\pi_{ES}$ and $\pi_{MS}$ approach $\infty$. This same behavior is observed for the metagrating analyzed in \cite{wirehk} (see Fig.~7 therein), where it is shown that as the radius of wires composing the metagrating vanish, $\pi_{MS}\rightarrow\infty$ and $\chi_{(ES,MS)}\rightarrow 0$.
Note that for circular apertures, the maximum possible fill-factor is $a/p=1/2$. Although in this limit the circular apertures do touch, the conducting screen does not vanish. As a result, $\chi_{(ES, MS)}$ do not approach zero and and $\chi_{MS}$ and $\pi_{(ES, MS)}$ do not approach $\infty$. In the $a/p\rightarrow 1/2$ limit for circular apertures, all the surface parameters approach finite, non-zero, values.

\section{Discussion and Conclusion}

We have investigated the interaction of electromagnetic fields with a symmetric metascreen. The surface parameters (the effective electrical and magnetic surface susceptibilities and surface porosities) that appear explicitly in the GSTCs are uniquely defined, and as such serve as the physical quantities that most appropriately characterize the metascreen. The effective surface parameters for any given metascreen together with the GSTCs given in (\ref{gstc1b}) and (\ref{gstc2b}) are all that are required to model its interaction with an EM field.

In this paper, we use the GSTCs to derive the reflection and transmission coefficients for a metascreen, which are expressed in terms of the electrical and magnetic surface parameters (surface susceptibilities and surface porosities). It is interesting to note that for TE-polarization, only two surface parameters ($\chi_{MS}^{xx}$ and $\pi_{MS}^{xx}$) are needed to fully characterize a metascreen which is in contrast to a metafilm, where three different surface parameters are needed to characterize a metafilm \cite{hk3}, \cite{hk2}, \cite{awpl} and \cite{metafilmemc}. On the other hand, for TM-polarization, four surface parameters ($\chi_{MS}^{zz}$, $\pi_{MS}^{zz}$, $\chi_{ES}^{yy}$, and $\pi_{ES}^{yy}$) are need to fully characterize a metascreen which is in contrast to a metafilm, where three different surface parameters are needed to characterize a metafilm for this polarization \cite{hk3}, \cite{hk2}, \cite{awpl} and \cite{metafilmemc}.

We show that knowing the reflection and transmission coefficients (obtained either from measurements or numerical simulations) of a metascreen, we can develop retrieval techniques for determining the surface susceptibilities and surface porosities, and hence a method for uniquely characterizing the metascreen. We demonstrate this retrieval approach by showing results for metascreens composed of five different apertures (circular holes, square holes, crosses, and slots, and square apertures filled with a high-contrast dielectric). We show that internal resonances associated with the material filling the apertures can give rise to interesting reflection and transmission behavior, i.e., enhanced transmission when compared to apertures with no filling. We also discuss behavior of the surface parameters in the two extreme limits of the fill-factor.

We have considered only the case where
the apertures and the array lattice have sufficient symmetry such that the magnetic surface susceptibilities and surface porosities of the resultant metascreen are diagonal (i.e., no cross-polarization terms). The presence of these more general GSTCs (cross-polarization or off-diagonal magnetic surface parameters) results in coupling between a TE and TM fields (i.~e., a TE polarized field will generate TM fields and vice-versa), and the reflection and transmission
coefficients for this more general case is the topic of a separate publication \cite{aniso-metascreen}.

\end{document}